\documentclass[journal]{IEEEtran}
\usepackage{mathrsfs}
\usepackage{amsmath}
\usepackage{amsthm}
\usepackage{cases}
\usepackage{tabularx}
\usepackage{diagbox}
\usepackage{booktabs}
\usepackage{amssymb}
\usepackage{graphicx}
\usepackage{times}
\usepackage{latexsym}
\usepackage{graphicx}
\usepackage{subfigure}
\usepackage{cite}
\usepackage{balance}
\usepackage{multirow}
\usepackage{algorithm}
\usepackage{color}
\usepackage{algorithmicx}
\usepackage{algpseudocode}
\usepackage{url}
\usepackage[justification=centering,labelsep=period]{caption}
\usepackage[table]{xcolor}
\newlength\savedwidth

                   \usepackage{array}
\pdfoutput=1

\begin{document}
\title{Cross-layer based intermittent jamming schemes for securing energy-constraint networks}
\author{Qinghe Gao\IEEEauthorrefmark{1}, Yan Huo, Tao Jing, Liran Ma, Jin Qian
\thanks{Qinghe Gao (Corresponding author) is with the School of Electronic and Information Engineering, Beijing Jiaotong University, Beijing, China, Email:
\texttt{qhgao@bjtu.edu.cn}}
\thanks{Yan Huo and Tao Jing are with the School of Electronic and Information Engineering, Beijing Jiaotong University, Beijing, China, Email: \texttt{\{yhuo,tjing\}@bjtu.edu.cn}}
\thanks{Liran Ma is with the Department of Computer Science, Texas Christian
University, Fort Worth, TX, USA, Email: \texttt{l.ma@tcu.edu}}
 \thanks{Jin Qian with the College of Computer Science \& Technology, Taizhou
University, Jiangsu, China, Email:
 \texttt{qianjin@tzu.edu.cn}}
}
\maketitle

\begin{abstract}
The Internet-of-Things (IoT) emerges as a paradigm to achieve ubiquitous connectivity via wireless communications between kinds of physical objects. Due to the wireless broadcasting nature and the energy constraint of physical objects, concerns on IoT security have triggered research on cooperative jamming based physical layer security. With the help of a cooperative jammer, existing solutions can effectively fight against eavesdroppers. However, these schemes are of high energy cost due to continuously transmitting jamming signals.
To reduce the energy consumption, we propose a new idea of intermittent jamming and design five specific intermittent jamming schemes (IJSs). By taking the transmit frame formate into account, we optimize these IJSs from three aspects, including the jamming power, the jamming method, and the jamming positions. Then we analyze the applicability of the proposed IJSs according to different requirements on the synchronization, the available jamming energy and the jamming power constraints. Extensive MATLAB experiments are conducted on the basis of the WLAN Toolbox, which demonstrate the proposed IJSs can effectively degrade the reception of the eavesdropper and outperform the widespread continuous jamming scheme (CJS) when the available jamming energy is limited.
\end{abstract}


\section{Introduction}
\label{sec:intro}

\IEEEPARstart{T}{he}  $\mathbf{a}$  development of the Internet of Things (IoT) has triggered research on secure and green communications\cite{2018:IoT:SecEne,2019:IoT:Secure,2015:IoT:PLS}. On one hand, the communication in IoT usually involves with much sensitive information, such as patients' private record\cite{2019:IoT:Health}, confidential sensor data \cite{2019:IoT:SensData}, and important financial messages \cite{2019:Access:IoTBank}, etc, which makes the security problem a growing concern\cite{2019:COMST:IoTSec,2019:Access:IoTSec} and security solutions more important. On the other hand, IoT terminal devices are often lightweight and battery-enabled, which requires security solutions to be of low complexity and of high energy efficiency. Recently, physical layer security (PLS) has emerged as a promising method to protect legitimated communication from eavesdropping in IoT networks \cite{2018:IoTJ:CJPLS,2018:CIOT:IoTEave}. This method takes advantage of mutual interference between users in physical layer to degrade the eavesdropping channel, which is much less complex than the conversional network layer cryptography.

To be specific, cooperative jamming based PLS schemes are widely applied. These schemes leveraged the cooperation between users to inject jamming signals into the eavesdropping channel so as to degrade the quality of the reception at  eavesdroppers without influencing the decoding at legitimated destinations \cite{2019:TWC:JamEave,2018:TVT:CJAF,2014:TNET:JamEav}. For the purpose of improving the security performance, these schemes usually employ continuous jamming signals to maximize the secrecy capacity (SC) \cite{2017:Gao:IET,2016:Gao:Globecom,2019:Wen:TVT} or minimize the probability of secrecy outage (SOP)\cite{2017:TVT:JamSOP,2018:TWC:JamSOP}.
However, the continuous jamming leads to high energy cost of the cooperative users. Many efforts have been made to solve this problem by improving the secrecy energy efficiency (SEE) \cite{2018:Wen:SEE} via optimizing the jamming power allocation. Furtheremore, much continuous jamming signals produce environmental noise pollution. In order to reduce unnecessary noise and enhance the jamming energy efficiency, we propose a novel idea of intermittent jamming to secure energy-constraint networks.

Under an intermittent jamming scheme (IJS), the jammer alternates between transmitting jamming signals to guarantee the security requirement and being silent to save energy and reduce noise pollution. We have analyzed the feasibility  of the IJS in our prior work \cite{2019:GAO:TCOM}, where an optimal jamming proportion has been obtained. However, different parts of the transmit signal has different importance for decoding the legitimate messages. Thus, the design of an energy-efficient IJS still faces many challenges, such as i) when is the best time for the jammers to transmit jamming signals, ii) how long time each jamming duration should be, and iii)  what is the alternate frequency of the jammers between jamming and sleeping, etc.

In this paper, we aim to design  effective IJSs to disturb the reception of the eavesdropper by using a limited amount jamming energy. We build a two-layer system model, including the IoT-based device layer and the WLAN-based control layer. In order to realize the secure communication between energy-constraint IoT devices, we design five IJSs
based on WLAN frame formats in the control layer, and optimize the performance in terms of symbol-error-rate-energy-efficiency (SEREE) from three aspects. First, the jamming results influenced by the jamming power follows an additive effect and a ceiling effect. We obtain the minimum jamming power to produce an symbol error based on these two effects. Second, we discover a multiplying effect where  multiple jamming pulses can bring more symbol errors.  By using this effect, we design multiple jamming methods to enhance the jamming influence on the eavesdropper. Third, we develop the cascading effect of jamming the preamble field of the transmit frames where once the critical part is interfered, the whole frame is wrongly decoded. According to this strength, we propose the preamble-targeted IJS to effectively protect the legitimate message from eavesdropping.
We further discuss the applicability of each IJS according to their requirements on the synchronization. Finally, we verify the effectiveness of the proposed IJSs through extensive simulation experiments under different available jamming energy.

The main contribution of our work are summarized as follow.
\begin{itemize}
  \item We propose a novel IJS to improve the security performance under the energy constraint based on a two-layer system model and creatively take the transmission frame format into account for the design of detailed IJSs.
  \item We conduct detailed analyses of the influence on the jamming effect from three aspects: i) the jamming power, ii) the jamming method, and iii) the jamming positions, and optimize the IJS based on their characteristics of the additive effect, the multiplying effect and the cascading effect, respectively.
  \item We conduct extensive experiments on the SEREE performance of the eavesdropper
      under different available jamming energy. Simulation results verify our analyses and demonstrate that our proposed IJS can outperform the continuous jamming scheme (CJS) under low energy constraints.
\end{itemize}

The rest of the paper is organized as follows. We give the review of related works in Section \ref{sec:relatedwork}.  Then preliminaries of our
system model and a brief review of the transmission protocol
is given in Section \ref{sec:model}. The optimization problem is described
in \ref{sec:problem}, followed by scheme designs and discussions in Section \ref{sec:scheme}. Experiment simulations and analyses are given in Section \ref{sec:simulation}. Last, we conclude the paper in section \ref{sec:conclusion}.

\section{Related Work}
\label{sec:relatedwork}

The jamming technique was at first utilized to transmit intentional interference signals to hinder legitimate wireless communications \cite{2011:DosJam:SURV,2005:MobiHoc:RanReJam,2011:JamAttack:MILCOM, 2014:ReactiveCR:ACM,2016:ReactiveJam:TDSC}.
In this case, the jammer is the adversary of the legitimate users and is called a malicious attacker. With the development of the cooperative communication, instead of adversaries, the jammers have been also employed as helpers. On one hand, they are utilized as helpful defenders to disable the unauthorized transmission in order to prevent illegal information from being transmitted to legitimate receivers \cite{2013:SSP:JamDefense,2018:TVT:ReaJam}. On the other hand, an increasing number of researchers utilize the jammers in the physical layer security \cite{2018:TVT:SEE,2019:TCOMM:SEE,2019:COMST:Survey}. They are called as friendly jammers, which is the case we investigate in this paper.

Under a typical cooperative jamming based physical layer security scenario consisted of a legitimate transmitter, a legitimate receiver, an eavesdropper and a jammer, the friendly jammer is employed to transmit friendly jamming signals to protect the legitimate signal from eavesdropping. The requirement for the jamming signals is to degrade the eavesdropping channel with limited influence or without any influence on the legitimate channel. A large of amount work have studied the improvement of the secrecy capacity with completely known channel state information (CSI)\cite{2015:TPDS:JamSC,2017:TVT:JamSR}, the minimization of the secrecy outage probability with uncomplete CSI\cite{2017:TVT:JamSOP,2018:TWC:JamSOP} from the aspects of jammer selection \cite{2016:Gao:Globecom,2017:Gao:IET}, jamming power optimization \cite{2013:JSAC:JamPower}, and jamming beamforming design \cite{2013:TIFS:JamBeam}, etc. With the increasing concern on the green communication, researchers investigated the energy efficiency of these jamming schemes. For example, the authors proposed a resource allocation algorithm to maximize the energy efficiency of the secondary user while guaranteeing a minimum secrecy rate for the primary user in a cognitive radio scenario \cite{2015:LWC:Gabry}. With statistical CSI of eavesdroppers, Hu et al. investigated the secrecy energy efficiency under the SOP constraint\cite{2018:TVT:SEE}. The authors improved the SEE under a multiple-input-multiple-output multiple-antenna eavesdropper model by optimizing the transmitted beamforming vectors \cite{2019:TCOMM:SEE}.

However, existing energy efficient cooperative jamming schemes are based on the assumption that the jammer is of enough available jamming energy to provide continuous jamming signals during the legitimate transmission. To save jamming energy, Allouche \textit{et al.} \cite{2017:Yair:TimeJam} proposed temporal jammers that produced noise for some portion of the time to affect a subset of the bits in a received message of the eavesdropper. By jointly optimizing multiple  jammers' jamming time and positions in space, they constructed a secure zone to prevent eavesdroppers from receiving correct legitimate messages. Motivated by this work, we investigated the feasibility of the IJS in the physical layer security in our prior work \cite{2019:GAO:TCOM}, which has demonstrated the advantages of the IJS over the continuous jamming schemes in terms of the bit error rate energy  efficiency under low available jamming energy.  As an extended work, we design detailed IJS schemes based on the transmission protocol in terms of the jamming durations and jamming intervals.

These designs can be referred to the works for the cases where the jammers are attackers or defenders. Taking the energy conservation into consideration,  Xu \textit{et al.} proposed a random jammer that periodically transmit jamming signals \cite{ 2005:MobiHoc:RanReJam}. By adjusting the distribution of the sleeping phase $t_s$ and jamming duration $t_j$, they achieved various levels of tradeoff between energy efficiency and jamming effectiveness. Researchers also proposed a reactive jammer that transmitted jamming signals reactively when it sensed activity on the channel \cite{2011:JamAttack:MILCOM,2014:ReactiveCR:ACM,2016:ReactiveJam:TDSC}. By this reactive method, adversaries can reduce the network throughput using little energy while minimizing the chances of being discovered.  Similarly, Jang \textit{et al.} proposed a reactive symbol-level jamming technique where the jammer sends random digital modulation symbols to the malicious receiver after it senses the transmission of the malicious  transmitter and detects the modulation scheme used at the malicious transmitter \cite{2018:TVT:ReaJam}. In this way, the malicious links are disrupted by the reactive jamming.

However, the designs of the IJS for the physical layer security face more challenges because the transmitted message cannot be protected during the non-jamming durations. In this paper, we propose to inject jamming signals to destroy the critical bits in the transmitted message so that effectively prevent the eavesdropper from correctly decoding its received messages. Based on the communication protocol, we design five IJSs to guarantee different levels of security under energy constraint networks.

\section{Preliminaries}
\label{sec:model}
\subsection{Network Model}

\begin{figure}
  \centering
  \includegraphics[scale=0.4]{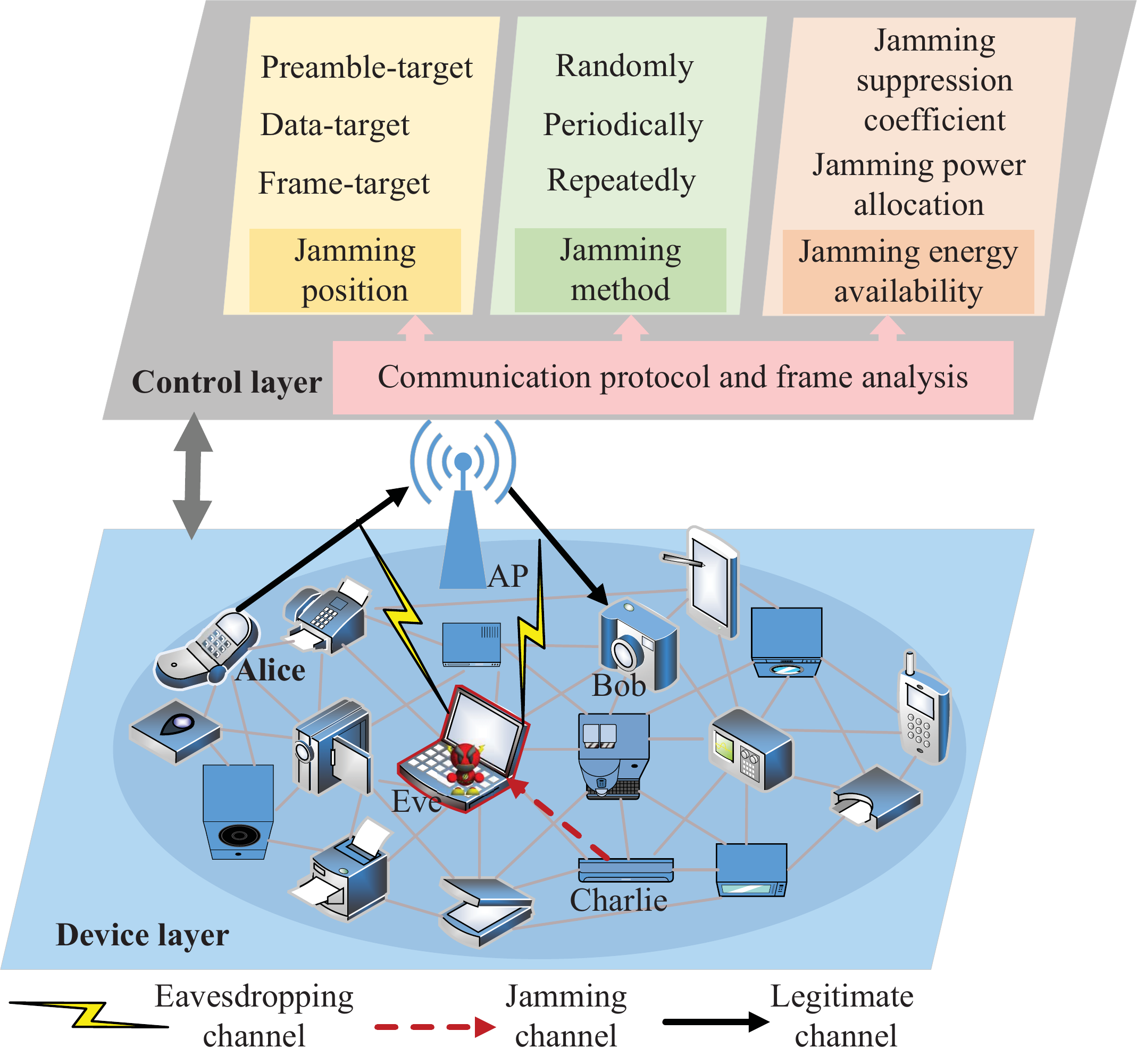}\\
  \caption{System model}\label{fig:model}
\end{figure}

\begin{figure*}[htbp]
  \centering
  \includegraphics[scale=0.5]{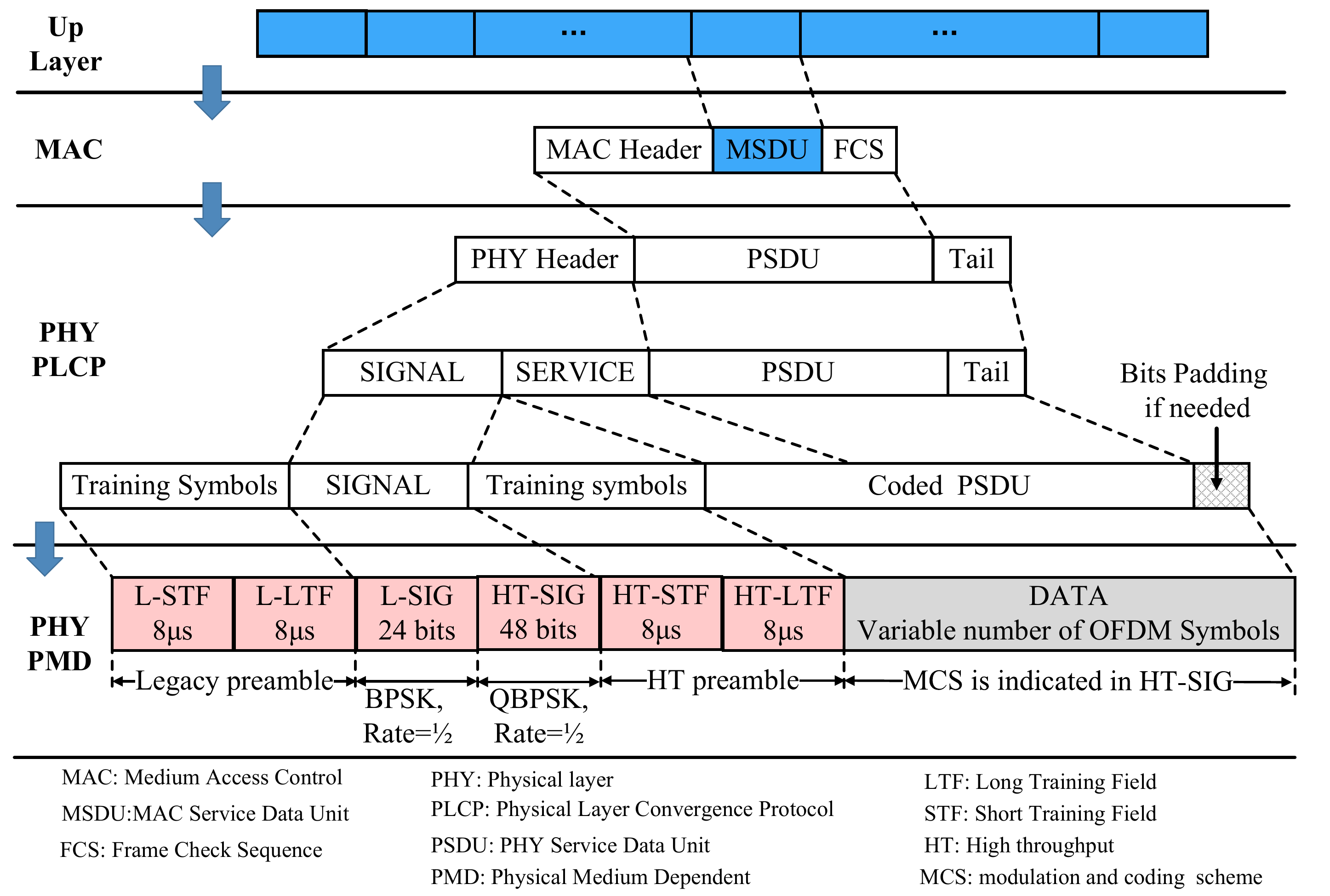}\\
  \caption{Transmission protocol and frame generation}\label{fig:phyframe}
\end{figure*}

 In the WLAN-based IoT, shown as the device layer in Fig.~\ref{fig:model}, any two IoT devices can communicate with each other in the way of wireless communication. Specifically, when a source  device (Alice) wants to transmit message to a destination device (Bob), it first sends the message to the access point (AP), then AP forwards the message to Bob. The communication channels from Alice to AP and from AP to Bob are called legitimate channels, in which the signals being transmitted are named as legitimate signals. During the transmission process, the legitimate signals can also be overheared by other non-target devices. Generally, these devices shall discard the signals that are not directed to them. While an intentional eavesdropper (Eve) may exist to intercept information from the signals transmitted by Alice or AP. The channels from Alice to Eve and from AP to Eve are called as the eavesdropping channels. Degrading the quality of the eavesdropping channels is an effective way to combat the eavesdropping. To this end, another device can be employed as a friendly jammer (Charlie) to transmit jamming signals without influencing the receptions of AP and Bob so as to preserve the secrecy of the legitimate information.

Traditionally, the jamming signals completely cover the  transmission process. The performance of this continuous jamming scheme (CJS) is in proportion to the jamming power. However, the jammer, as an IoT device, is usually of low energy that is not enough to provide satisfied security performance. Thus, we consider an IJS where the jammer can alternate between transmitting jamming signals to interfere with the eavesdropper and keeping sleep to save energy. In this way, the limited jamming energy is efficiently utilized to guarantee that Bob can receive correct messages and Eve receives a version with as many errors as possible. In the following sections, we design the detailed jamming pattern on the basis of the communication protocol and the transmission frame format, shown as the control layer in Fig.~\ref{fig:model}. Specifically, we determine when the jamming should start and end by analyzing the influences of different jamming positions on the transmitted signal, design jamming methods to decide how often the jamming injections occur, and optimize how much the jamming power should be under the analysis on the jamming energy availability.

\subsection{Transmission Protocol and Frame Format}\label{sec:protocol}
In this paper, we employ the configurations based on IEEE 802.11n standard for higher throughput communication \cite{2009:80211Std:HT}, which supports the multiple-input-multiple-output (MIMO) in the physical layer. This can be utilized to cancel the jamming noise at legitimate receivers by beamforming. The orthogonal frequency division multiplexing (OFDM) physical layer specification is employed and the transmit signal format is of the high-throughput (HT) type. Under the channel bandwidth of 20MHz, the transmission protocol and the frame generation process are shown in Fig.~\ref{fig:phyframe}.

We focus on the medium access control (MAC) and the physical layer (PHY) protocol. In the MAC layer, the user's message coming from the higher layers are divided into multiple segments to accomodate the length of one MAC service data unit (MSDU). Each segment is enclosed into the MSDU and a MAC frame is generated by adding MAC header and frame check sequence (FCS) respectively in front of and in the end of the MSDU. Then, the whole MAC frame passes into the PHY layer. The PHY layer consists of the physical layer coverage procedure (PLCP) sublayer and the physical medium dependent (PMD) sublayer. The PLCP sublayer simplifies the PHY service interface to the MAC services. The PMD sublayer provides a means to send and receive data between devices.

In the PLCP sublayer, the MAC frame is taken as the physical layer service unit (PSDU). Adding physical layer (PHY) header in front of the PSDU and tail bits following the PSDU, the physical layer frame is generated. The PHY header consists of SIGNAL and SERVICE, where the SIGNAL contains the legacy SIGNAL (L-SIG) and HT SIGNAL (HT-SIG). The L-SIG contains rate, length, and parity information. The HT-SIG carries information used to decode the HT packet, including the modulation and coding scheme (MCS), the packet length, the FEC coding type, etc. The HT-SIG is composed of two symbols, HT-$\mathrm{SIG_1}$ and HT-$\mathrm{SIG_2}$.  Each SIGNAL symbol is of 24 bits and coded in BPSK with coding rate of $r=1/2$, then transmitted by one OFDM symbol in  $4 \mu s$, that is $t_{LSIG}=4\mu s$ and $t_{HTSIG}=8 \mu s$.

During transmission, training symbols are inserted in front of the L-SIG and following the HT-SIG. Each training symbol consists the short training field (STF) and the long training field (LTF). The STF and LTF in front of the L-SIG are called as the legacy STF (L-STF) and the legacy LTF (L-LTF). The L-STF is of time length as $t_{LSTF}=4\mu s$ and is used for the start-of-packet detection and the coarse frequency correction due to its good correlation properties. The L-LTF is also of time length as $t_{LLTF}=4\mu s$ and is used for channel estimation, fine frequency offset estimation, and fine symbol timing offset estimation.
The L-STF, L-LTF and the L-SIG construct the legacy preamble.

The STF and LTF following the HT-SIG are named as HT STF (HT-STF) and HT LTF (HT-LTF). The HT-STF is used to improve automatic gain control estimation for a MIMO system and the HT-LTF are necessary for demodulation of the following data field.
The HT-SIG, HT-STF and HT-LTF make up the HT preamble. Both HT-STF and HT-LTF are of time length as $t_{HTSTF}=t_{HTLTF}=4 \mu s$. 

The SERVICE bits, the PSDU, and the tail bits constitute the data field together with padding bits of zeros. The data field is modulated according to the rate and the MCS that are indicated in the SIGNAL parts, then converted into OFDM symbols. The number of data bits in an OFDM symbol, $L_{DBPS}$, is also determined by the rate of SIGNAL part. The total length of the data field shall be a multiple of $L_{DBPS}$, that is, the total number of OFDM symbols for the data field, denoted as  $N_{sym}$, is computed as $N_{sym}=\lceil \frac{16+8*LENGTH+6}{L_{DBPS}}\rceil$, where $\lceil\cdot\rceil$ represents the ceiling function, LENGTH is indicated in the L-SIG and represents the number of bytes for the PSDU.  Therefore, the number of the padding zero bits is $L_{pad}=N_{sym}*L_{DBPS}-(16+8*LENGTH+6)$. Finally, the total length of the data field is $t_{DATA}=(N_{sym}*4)\mu s$.

In the PMD layer, the transmission of a complete frame is appending the OFDM symbols one by one after the preamble. For the transmission of multiple frames, there is an idle time between two frames. The time length for a frame is computed as $t_{frame}=t_{preamble}+t_{DATA}=(36+N_{sym}*4)\mu s$. The idle time between each frame , $t_{idle}$, can be adjusted according to the channel bandwidth.

\section{Problem Formulation}
\label{sec:problem}

 In this section, we assume that the message that is transmitted from Alice to Bob is of $L_{message}$ bytes. According to the transmission protocol, this message will be enclosed to the payload of MAC frames. If the maximum length of the MSDU is denoted as $L_{msdu}$ bytes, the total number of the frames needed to transmit the whole message is calculated as
   \begin{equation}\label{eq:pktnum}
     N_{frame}=\lceil\frac{L_{message}}{L_{msdu}}\rceil,
 \end{equation}
 with the last MSDU padded with zeros of length $L_{pad}^{message}=L_{msdu}N_{frame}-L_{message}$ (bytes) following the message bits. Thus, the time length of the legitimate signal is computed as
 \begin{equation}\label{eq:timesignal}
   t_{signal}= N_{frame}*(t_{frame}+t_{idle}).
 \end{equation}

The complex envelope of the transmitted signal can be modeled  as  $s(t)$ and $s_j(t)$ for the legitimated signal and the jamming signal, respectively. The channel gain between users can be denoted as $h_{ab}$ with $a \in \{A,J\}$ for Alice and Charlie and $b \in \{B, E\}$ for Bob and Eve, respectively.
Under the sampling rate $f_s$ (e.g., 20MHz), we can obtain the discrete version of the received data at Bob and Eve as follows,
\begin{eqnarray}\label{eq:model}
  r_B(k) = h_{AB}s(k)+h_{JB}\beta(k)s_j(k)+n_B(k), 1\leq k\leq N_s, \\
  r_E(k) = h_{AE}s(k)+h_{JE}\beta(k)s_j(k)+n_E(k), 1\leq k\leq N_s,
\end{eqnarray}
where $n(k)$ is the discrete zero-mean additive white Gaussian noise (AWGN) with variance as $N_0$. $N_s$ is the total sampling number of the whole transmission stream and is calculated as $ N_s = N_{frame}t_{frame}f_s$. $\beta(k)$ is an indicator of the presence of the jamming signal in the $k^{th}$ sampling interval. According to different jamming schemes, $\beta(k)$ is of different values. Specifically, in the CJS, all $\beta(k)$ equals 1. For the IJS, $\beta(k)$ can be either deterministic value or a sequence of independently discrete random variables taking values in the set $\{0,1\}$. Therefore, the intermittent jamming is similar to the pulse jamming \cite{2011:JammingTech}. The width of the pulse represents the jamming duration, the duty cycle of the pulse determines the jamming interval where there is no jamming signal transmitted.

The actual jamming power in the IJS can be calculated as
$ P_j=\frac{\Sigma_{k=1}^{N_s}\beta(k)|s_j(k)|^2}{\Sigma_{k=1}^{N_s} \beta(k)}$,
 where $\Sigma_{k=1}^{N_s}\beta(k)$ determines the width of the jamming pulse and $|s_j(k)|^2$ represents the strength of the jamming pulse. According to the above subsection, the data field is modulated before converting into OFDM symbols, take the 64QAM with coding rate $r=3/4$ as an example, the symbol error rate (SER) received at Eve can be calculated as
 \begin{equation}\label{eq:SER}
   Pr=1-\left[1-\frac{3}{4}Q\left(\sqrt{ \frac{2}{7}\frac{|h_{AE}|^2E_b}{|h_{JE}|^2P_j}}\right)\right]^2,
 \end{equation}
 where $E_b$ is the energy for each bit in the legitimate signal and is calculated as $E_b=4*f_s|s(k)|^2/L_{DBPS}$, $Q(\cdot)$ is the Q-function.  Generally, for a jamming pulse with a certain width, a higher jamming power means a higher probability that bits influenced by this jamming pulse are wrongly decoded. This jamming effect influenced by the jamming power can be considered as an additive effect. However, the maximum value of the bit error probability is 0.5, the jamming power beyond a threshold does not raise the jamming effect, which can be called as a ceiling effect.

 This ceiling effect can be cancelled by weakening the strength of the jamming pulse and increasing the influence coverage of the jamming pulse. On one hand, the surplus jamming strength can be utilized to interfere with more samples. On the other hand, dividing a wide jamming pulse into multiple narrower pulses can enhance the jamming effect because the transmitted signal are M-ary modulated where a bit error means a symbol error. Thus, we need to design effective jamming methods to decide the width and the number of the jamming pulses. Usually, more jamming pulses means more errors, which can be seen as a multiplying effect. Since $L_{DBPS}$ bits are modulated and then converted into one OFDM symbol, we can obtain the number of the modulated symbols transmitted in each OFDM symbol as $N_{sym}^{coded}=\frac{L_{DBPS}/r}{\log_2M}$, where $r$ is the coding rate and $M$ is the modulation order, e.g., $M=6$ for the 64QAM. The number of the error symbols can be calculated as
 \begin{equation}\label{eq:numsymerr}
   N_{error}=\min\left\{\lambda \frac{\Sigma_{k=1}^{N_s}\beta(k)}{4f_s}N_{sym}^{coded}Pr,N_{sym}^{total}\right\},
 \end{equation}
where $\lambda$ is the multiply effect parameter, $\frac{\Sigma_{k=1}^{N_s}\beta(k)}{4f_s}N_{sym}^{coded}$ is the number of symbols that are covered by the jamming signals, $N_{sym}^{total}=N_{frame}N_{sym}N_{sym}^{coded}$ is the total number of the transmitted symbols for data fields, and  $Pr$ is given by \eqref{eq:SER}.

Furthermore, according to the transmission protocol, the preamble, especially the HT-LTF, is responsible for decoding the data field. Specifically, the estimation of the HT-LTF field provides the reference noise level for the decision of the decoding process. Once the HT-LTF is interfered by the jamming signals, the noise level is wrongly estimated, resulting in random decision on 0 or 1 with equal probability of 0.5. Thus, once the HT-LTF is jammed, almost all symbols are wrongly decoded, which can be considered as a cascading effect. However, eavesdroppers may bypass the preamble field to obtain the data field via some history values of the noise levels. Therefore, according to the different importance of different fields in each frame, we define a weight parameter as $w=1$ to indicate that the HT-LTF field is completely jammed, and $w=0$ for other cases.

 Jointly considering the influence of the jamming power, the jamming methods, and the jamming positions, the overall number of the symbol errors are denoted as
\begin{equation}\label{eq:HTLTFerror}
  N_{error}^{overall}=wN_{sym}^{total}+(1-w)N_{error}, w\in \{0,1\}.
\end{equation}
 Our goal is to find the effective IJS under a limited amount of jamming energy to prevent the eavesdropper correctly decoding the legitimate message by employing as little energy as possible under the guarantee that the legitimate receiver can decode the transmitted message in an error-free method. Thus, our IJS design problem can be formulated as
\begin{eqnarray}
\max_{\beta(k),k\in[1,N_s]}& \frac{N_{error}^{overall}}{N_{sym}^{total}E_J}\label{eq:objective}\\
s.t. &N_{error}^{overall}(h_{AB},h_{JB})=0.\label{eq:SERbob}\\
&E_J=\Sigma_{k=1}^{N_s}\beta(k)|s_j(k)|^2\leq E_J^{avail},\label{eq:AJE}\\
&\beta(k)=\{0,1\}, \label{eq:betak}\\
&w=\begin{cases}
1, \beta(k)=1, \forall k\in[32fs,36fs],\\
0,  \mathrm{otherwise},
\end{cases}\label{eq:weight}
\end{eqnarray}
The objective is to maximize the SER-energy-efficiency (SEREE) which is computed as the ratio between the overall SER at the Eve and the consumed jamming energy. The amount of the available jamming energy is denoted as $E_J^{avail}$ and the consumed jamming energy shall be not beyond $E_J^{avail}$ as shown by the constraint \eqref{eq:AJE}. Constraint \eqref{eq:SERbob} shows the requirements for error-free reception at Bob.

\section{Intermittent Jamming Scheme}
\label{sec:scheme}
In this section, we first give the details of the proposed IJS according to different categories. Then, we analyze the influence parameters of different schemes. Last, we discuss several critical points for our proposed schemes.

 As we know, the main purpose of the passive eavesdropper is to correctly obtain Alice's data rather than destroying the communication process between Alice and Bob. Thus, data frames, instead of protocol frames such as ACKs, are just what the eavesdropper most concerns about. According to the IEEE 802.11 standards, if a user receives a package with mistaken FCS, the package shall be discarded. On the basis of this rationality, we tend to trap  the eavesdropper  to discard its received package by effectively jamming critical bits in the transmitted message. Note that our scheme is still effective even if Eve is strong enough to reserve all received packages, because our goal is to make Eve received different versions of transmitted message under the guarantee that Bob receives correct messages.

 \subsection{IJS for Different Jamming Methods}\label{sec:jammingmethod}
\begin{figure}
  \centering
  \includegraphics[scale=0.3]{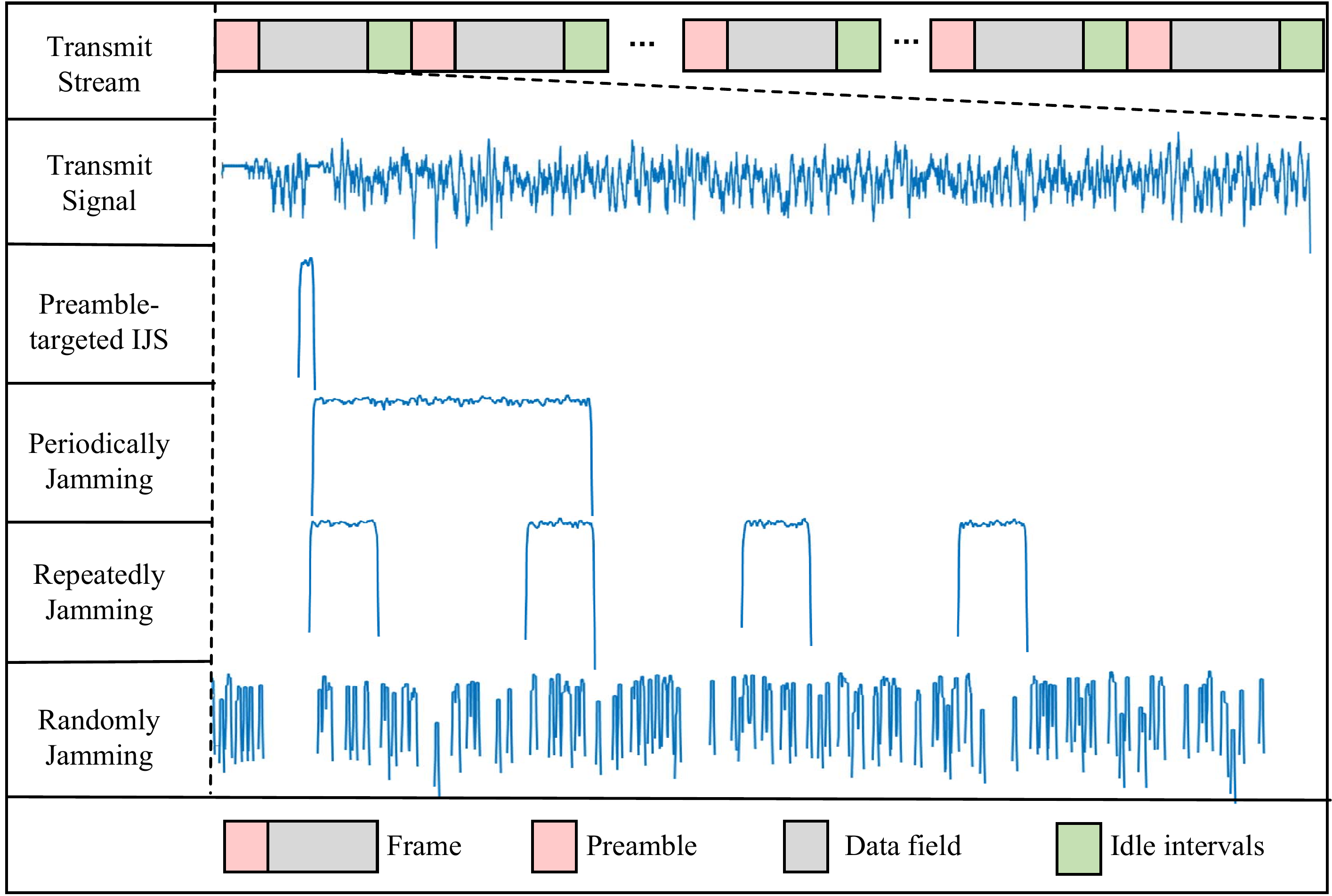}\\
  \caption{The jamming signals for IJS.}\label{fig:IJSslot}
\end{figure}

As we have mentioned, the jamming effect brought by different width of a jamming pulse and the total number of jamming pulses follows the multiplying effect. By utilizing this effect, we can adaptively employ different jamming methods to transmit jamming signals. According to the different characteristics of the jamming pulses, shown in the Fig.~\ref{fig:IJSslot}, we can conduct three jamming methods as follows.

\subsubsection{ Periodically jamming (PerJ)}
  In this method, only single one jamming pulse is transmitted during the transmission of each frame and is repeated for each frame. Specifically,  the jamming duration and the jamming interval, represented as $t_d$ and $t_v$, respectively,  satisfy the constrain that $t_d+t_v = t_{frame}$ for the case of the single jamming pulse.  Under this case, for the $n^{th}$ frame, we have
 \begin{equation}\label{eq:periodjamming}
    \beta_n(k) = \begin{cases}
              1, k\in[i,j],\\
              0, k\in[1,i)\cup(j,N_s^{frame}],
              \end{cases}
\end{equation}
where  $N_s^{frame}=t_{frame}*f_s$ is the number of samples for each frame.
This method concentrates the jamming power by block and block.

\subsubsection{Repeatedly jamming (RepJ)}
This method is similar to the periodically jamming, but there are multiple jamming pulses during the transmission of each frame and repeat the jamming pulses for following frames, that is, $N_{pulse}*(t_d+t_v)=t_{frame}$. This method can cause errors in  multiple blocks, which may conceal the critical information at a high probability. Under this case, for the $n^{th}$ frame, we have
   \begin{equation}\label{eq:repeatjamming}
    \beta_n(k) = \begin{cases}
              1, k\in[i_1,j_1]\cup[i_2,j_2]\cup \cdots \cup [i_{N_{pulse}},j_{N_{pulse}}],\\
              0, k\in[1,i_1)\cup(j_1,i_2)\cup \cdots \cup (j_{N_{pulse}},N_s^{frame}].
              \end{cases}
\end{equation}

 \subsubsection{ Randomly jamming (RanJ)}
   The errors caused by the above jamming methods with fixed jamming durations and jamming intervals may concentrate on a certain region that can be easily neglected by the eavesdropper. To obtain a disperse effect on the errors of the whole frame, we can employ a random jamming method, where $\beta(k)$ can randomly take value from $\{0,1\}$. Under this case, the jamming durations and intervals are dynamic and diverse such as
   \begin{equation}\label{eq:randomjamming}
    \beta_n(k) = \begin{cases}
              1, k\in \mathcal{K}, \\
              0, k \in \tilde{\mathcal{K}},
              \end{cases}
\end{equation}
where $\mathcal{K}$ is a subset of $[1,N_s]$ with each elements randomly selected from $[1,N_s]$, and $\tilde{\mathcal{K}}$ is the complementary set of  $\mathcal{K}$ with $\tilde{\mathcal{K}}\cup\mathcal{K}=[1,N_s]$ and $\tilde{\mathcal{K}}\cap\mathcal{K}=\phi$.

 \subsection{IJS for Different Jamming Positions}
 Among the received signal, the eavesdropper is most interested in the parts that correspond to the legitimate information. Thus, an intuitive method is to disturb the data parts. However, the decoding of the data part is based on the channel estimation results obtained from the preamble. Therefore, we classify our proposed IJS into three categories according to the jamming positions.

\subsubsection{ Preamble-targeted (PT) IJS}
  As we introduced in section \ref{sec:protocol}, the preamble of the frame is used for channel estimation. The channel estimation results including the estimation about the noise level and channel gains are employed into the demodulation of the data part. We propose injecting interference to this part so as to change the preamble and increase the estimated noise level. Then the polluted results cause mistakes to the decoding of the data part, which would be like randomly guessing the initial signal. It can degrade the eavesdropper effectively via a narrow jamming pulse by setting $t_d$ of the PerJ method as $t_d=t_{HTLTF}$, that is,
\begin{equation}\label{eq:preamblejamming}
    \mathbf{\beta}_n(k)= \begin{cases}
              1, k\in[32f_s,36f_s],\\
              0, k\in[1,32f_s)\cup(36f_s,N_s^{frame}].
              \end{cases}
\end{equation}

\subsubsection{Data-targeted (DT) IJS}
  Since the eavesdropper most concerns the data part, a smart one may be able to bypass the preamble test and harvest the critical data directly. To prevent this case from happening, we can focus on disturbing the data transmission. The injected noise changes the complex envelop of the received signal, then cause errors in the decoding results. Once the number of errors exceeds a certain value, the eavesdropper cannot guess the true version of the legitimate information.  Since the data field is of longer transmission than the preamble field, all of the three jamming methods can be employed to achieve good jamming effect.
\begin{equation}\label{eq:datajamming}
    \beta_n(k) = \begin{cases}
              1, k\in[i,j]\\
              0, k\in[1,i)\cup(j,N_s^{pkt}]
              \end{cases}, 36f_s<i,j<N_s^{pkt}.
\end{equation}
\subsubsection{ Frame-targeted (FT) IJS }
  Considering the advantages and disadvantages of the above two cases, we employ the frame-targeted scheme, where both the preamble and the data field in each frame are disturbed. The random jamming method is the most suitable one in this case because the preamble is included. This method can relax the requirements on the time synchronization and provide double damages to the reception of Eve.

\subsection{Applicability Discussion}
Jointly considering the jamming effect and the practicality of the jamming pulse, we obtain the matching table for different jamming positions and different jamming methods as follows. In this table, the matching results represent the applicability of each IJS. Each IJS demonstrates  the jamming signals shall be injected  into which positions by which method. For example, the PerJP  means jamming the preamble part in a periodical way. In summary, for jamming the preamble part, the periodically jamming is the best method, and if the target is jamming the whole frame, the best way is using the randomly jamming.
\begin{table}[h]
\centering
\caption{Proposed IJSs by matching the jamming positions and jamming methods.}
\begin{tabular}{c|c|c|c}
  \hline
  \diagbox[width=2.5cm]{Positions}{Methods} & Periodically  & Repeatedly  &  Randomly \\
  \hline
   Preamble & PerJPT & $-$     & $-$ \\
   Data  & PerJDT & RepJDT & RanJDT \\
   Frame    & $-$     & $-$     & RanJFT \\
  \hline
\end{tabular}
\end{table}

\subsection{Remarks}
\subsubsection{Synchronization}
As we can see from the above descriptions of IJSs, there is a demand on the synchronization between Alice and jammer to guarantee that the signals received at the eavesdropper are effectively disturbed. Different IJSs have different requirements on the synchronization, specifically, the order from the highest requirement to the lowest requirement is PerJPT$>$PerJDT$>$RepJDT$>$RanJDT$>$RanJFT. The PerJPT is of the highest demand of the synchronization because the preamble is at the fixed position of each frame and the critical preamble is of a very short length. In this case, the start position of the jamming signal needs strict and accurate control to guarantee precisely interfering with the HTLTF field and realize success jamming with the highest SEREE.

However, the influence factors on the synchronization\cite{2009:AHICI:Cho},  including the time needed for processing each packet, the transmitted time from transmitter to the receiver, the offset of the received signal, etc, are dependent on the hardware processing and the channel conditions. Thus, the latter four IJSs are more applicable in practical scenario with loose requirements on the synchronization.

\subsubsection{Signal Detection}
To realize the synchronization, the signal detection is necessary. On one hand, the jammer needs to know the starts of the legitimation then determines how to jam; on the other hand, the legitimate receiver needs to detect whether there are jamming signals to cancel the interference.

There are two types of signal detection: cross-correlation and energy detection. The former is useful for fine-grained detection of signals from a particular wireless standard  where their preambles are known \cite{2015:IWCNC:Syn}. This allows high-speed detection design and deterministic timing in operations. Since the legitimate signals are with known preamble, the jammer can utilize the cross-correlation detection to detect legitimate transmission. The latter provides coarse-grained detection to detect any kind of RF signals with or without known signal preamble. This kind of detection is suitable for Bob to detect whether there is jamming signals in its reception.  The energy detection scheme for the burst signals in \cite{2019:LSP:Oh} can be applied to detect the jamming pulses in the IJS.

\subsubsection{Interference-Cancelation at Bob}\label{sec:nojamforBob}
The main idea of the physical layer security is degrading the eavesdropper without influencing the legitimate receiver. The solutions lie on two aspects. On one hand, we can employ zero-beamforing to design jamming signals so that guarantee no jamming signals are received at Bob \cite{2018:TCOMM:Ma,2018:TIFS:Yu}. This method has been widely studied in existing works, which is applicable for the scenario where the jammer or Bob or both of them are equipped with multiple antennas.  On the other hand, for the scenario where both the transmitter and the receiver are equipped with the single antenna, the jamming signals can be cancelled at Bob by using the impulsive noise cancellation  methods \cite{2012:LSP:Yih}.

\section{Numerical Results and Analyses}
\label{sec:simulation}

In this section, we use the WLAN Toolbox in MATLAB to simulate the proposed IJSs.
\subsection{Experiment Settings}
\begin{figure}[h]
  \centering
\includegraphics[scale =0.25]{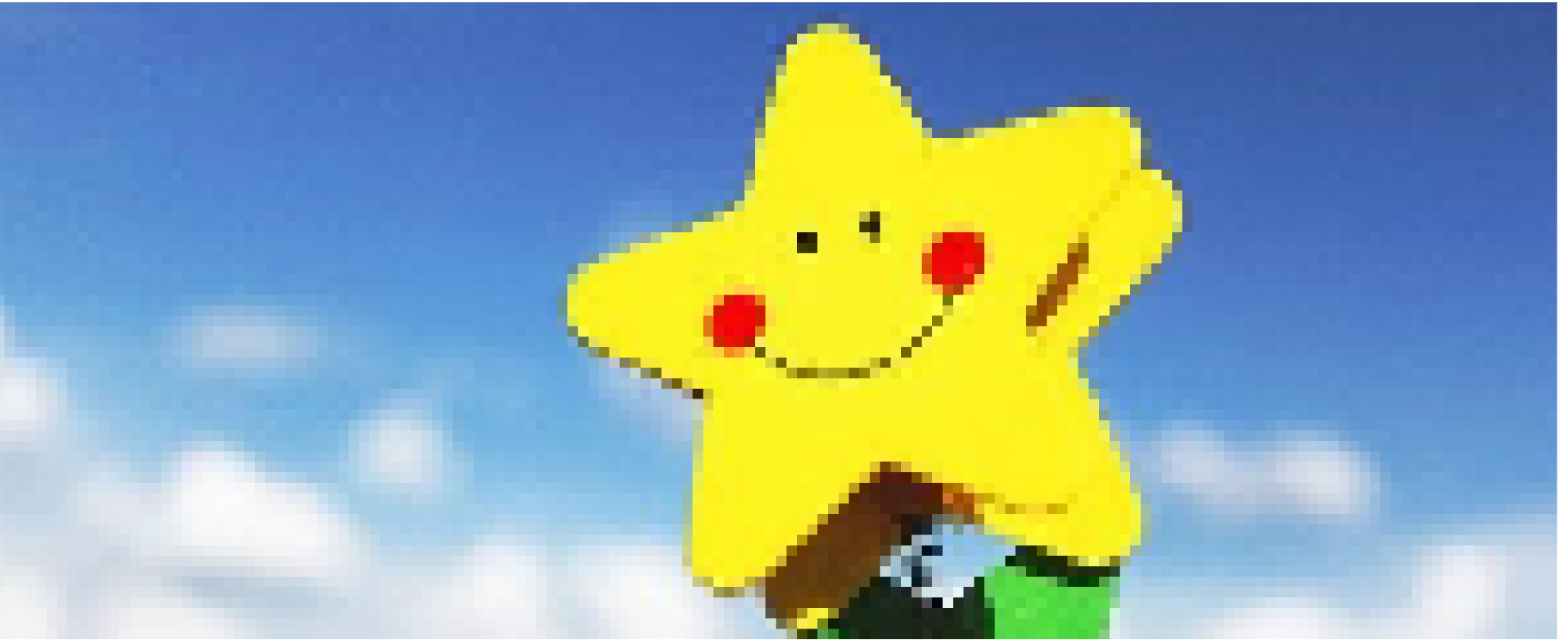}
  \caption{The correct picture}\label{fig:yuan}
\end{figure}

For the sake of clear illustration of the jamming effect, a picture transmitted from Alice is the message that needs to be protected. The picture is of the .jpg version with 61*150 pixels composed of the RGB color. Each pixel is represented as a uint8 number, thus, the length of the transmit message is 219.6MB, that is, $L_{message}=219600$ bytes. The picture Alice tends to transmit is shown as Fig.~\ref{fig:yuan}.

By using the WLAN Toolbox on the MATLAB R2019a platform, each node is of the 802.11n standard. Specifically, the channel model is the TGn channel, the MAC frame is of the HT format, the access protocol is OFDM, and the modulation method is 64-QAM.
\subsection{Measurement Parameters}
During the experiment, we fix the transmit power of Alice  and change the jamming energy (jamming power) and the jamming proportion under different jamming schemes. The influence parameters are listed as follows

\begin{itemize}
  \item Jamming Energy (JE): This parameter can be calculated as
   $JE=\Sigma_{k=1}^{N_s} \beta(k)|s_j(k)|^2$. Given a specific value of JE, we can select the jamming positions and the amplitude of the complex envelope  of the jamming signal. Specially, when the JE is lower, we should choose jamming the most critical positions such as the HTLTE.
  \item Jamming Proportion (JP): This parameter is defined as the ratio between the number of interfered positions and the total sampling number of the whole stream, which can be computed as $\rho =\frac{\Sigma_{k=1}^{N_s} \beta(k)}{N_s}$.
  \item Jamming-to-Signal Ratio (JSR): This parameter is defined as the average jamming power divided by the legitimate signal power, whose detailed expression is $JSR =\frac{\Sigma_{k=1}^{N_s}\beta(k)|s_j(k)|^2}{\Sigma_{k=1}^{N_s} \beta(k)}/\frac{\Sigma_{k=1}^{N_s}|s(k)|^2}{N_s}= \frac{JE}{\rho}\Sigma_{k=1}^{N_s}|s(k)|^2$. A larger JSR means more interference on the legitimate signal.
\end{itemize}

\subsection{Metric}
The SER at the Eve is used as the performance metric to measure the effectiveness of different jamming schemes. The premise is that we assume Bob can correctly receive the transmitted message as we have discussed in section \ref{sec:nojamforBob}. Thus, a higher BER of Eve means a better jamming effect.  The SER depends on two factors, one is the probability where an error occurs and another is the number for the potential samples being jammed. The former increases with the jamming power. The latter is proportional to the jamming proportion.

\begin{figure*}[htbp]
\centering
\subfigure[Jamming Signals]{
\begin{minipage}[b]{0.35\linewidth}
\includegraphics[width=1\linewidth,height=1.9cm]{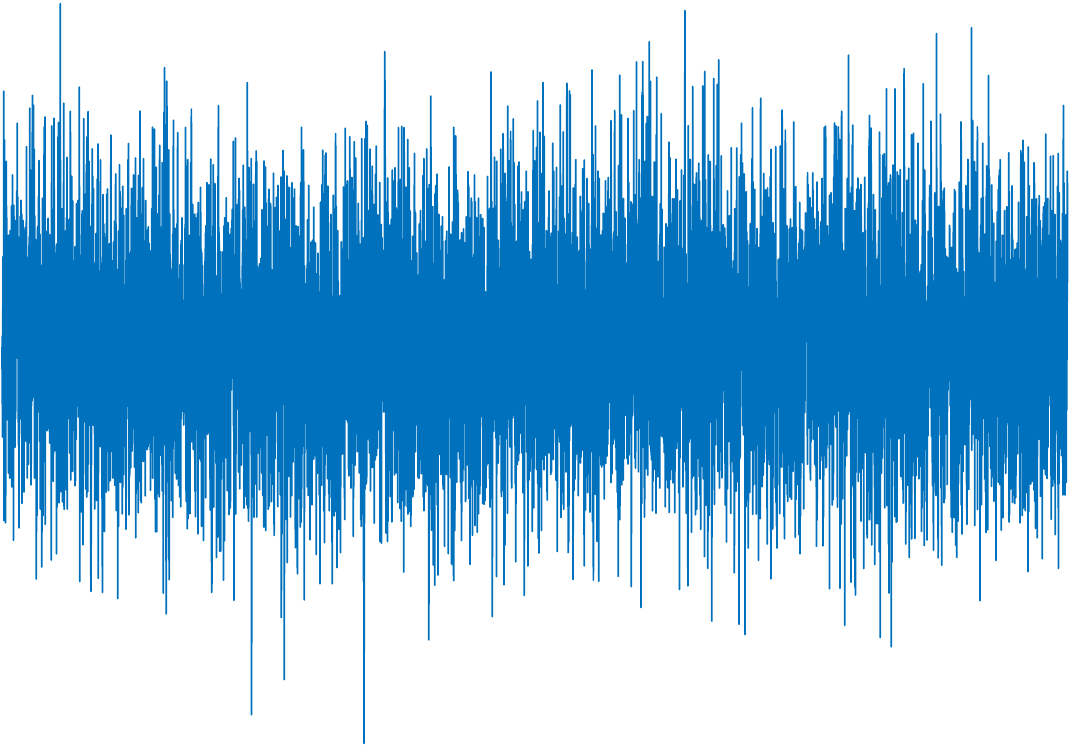}\vspace{4pt}
\includegraphics[width=1\linewidth,height=1.9cm]{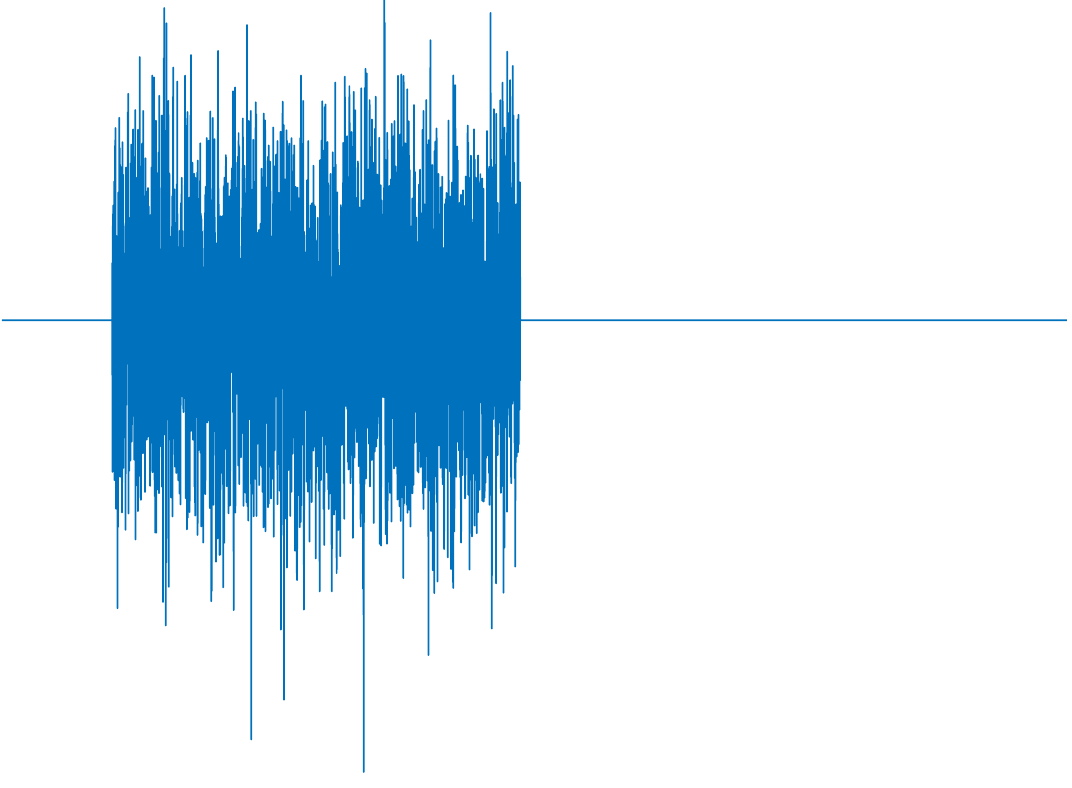}\vspace{4pt}
\includegraphics[width=1\linewidth,height=1.9cm]{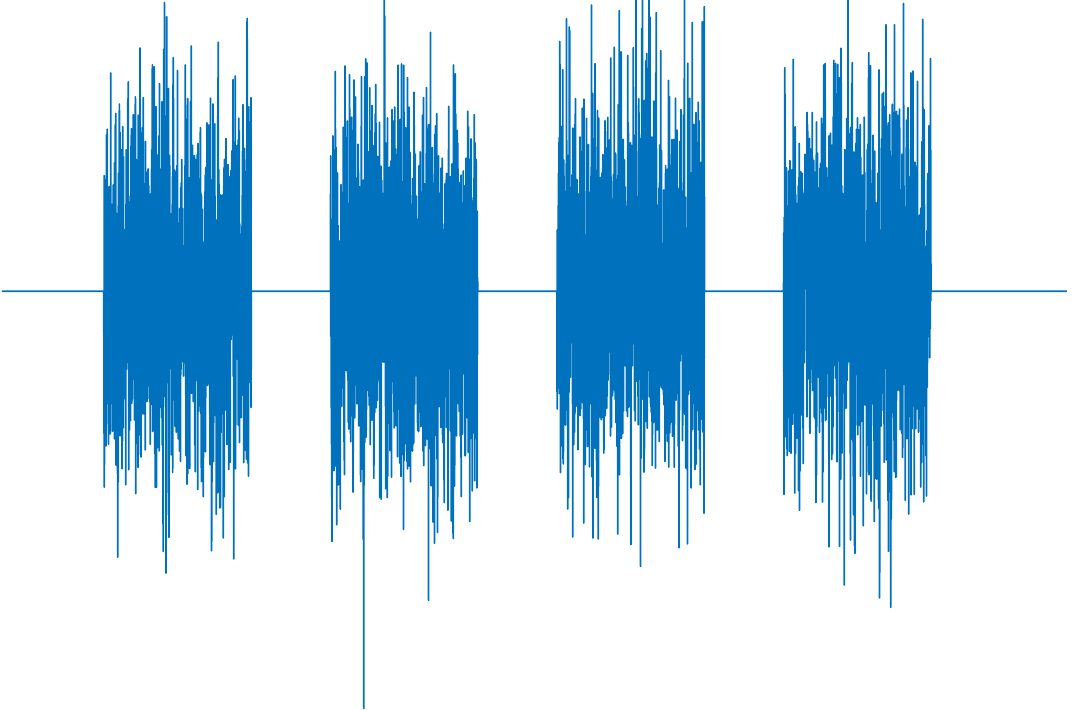}\vspace{4pt}
\includegraphics[width=1\linewidth,height=1.9cm]{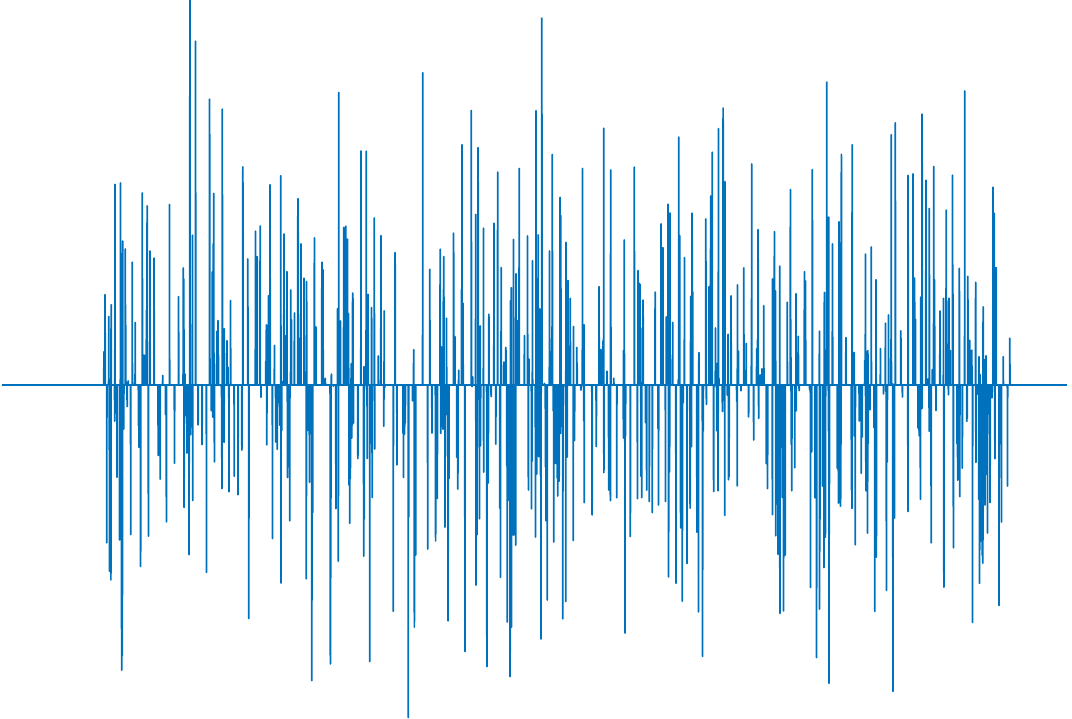}\vspace{4pt}
\includegraphics[width=1\linewidth,height=1.9cm]{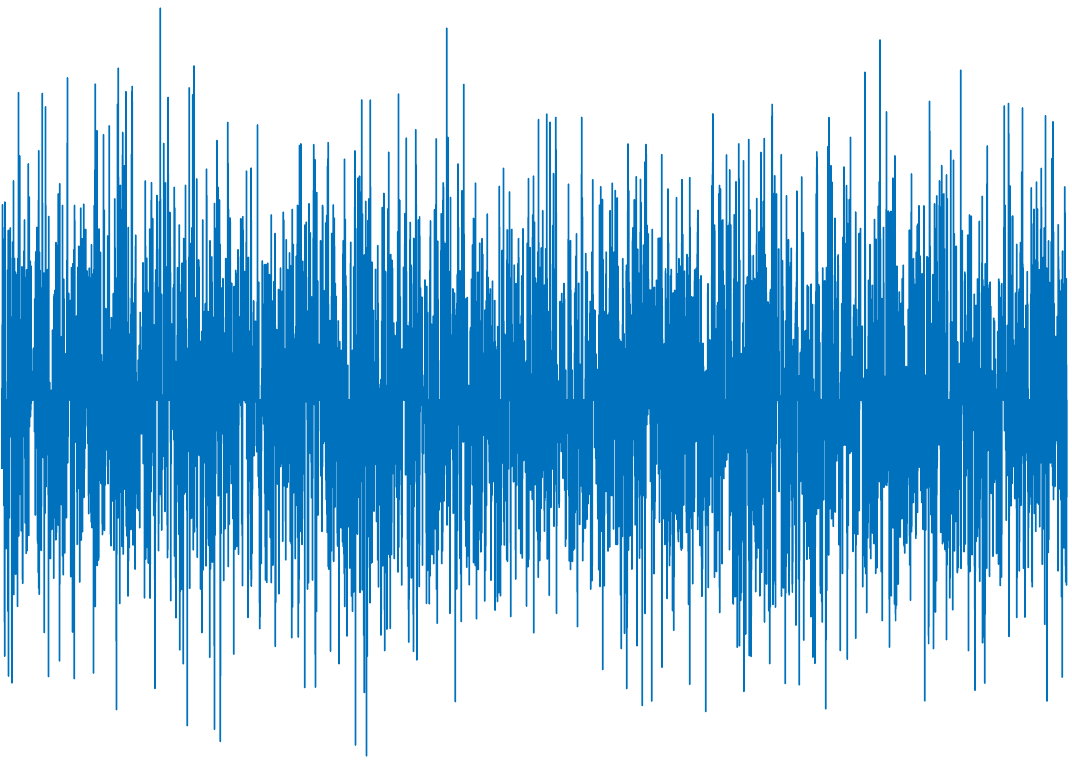}\vspace{4pt}
\includegraphics[width=1\linewidth,height=1.9cm]{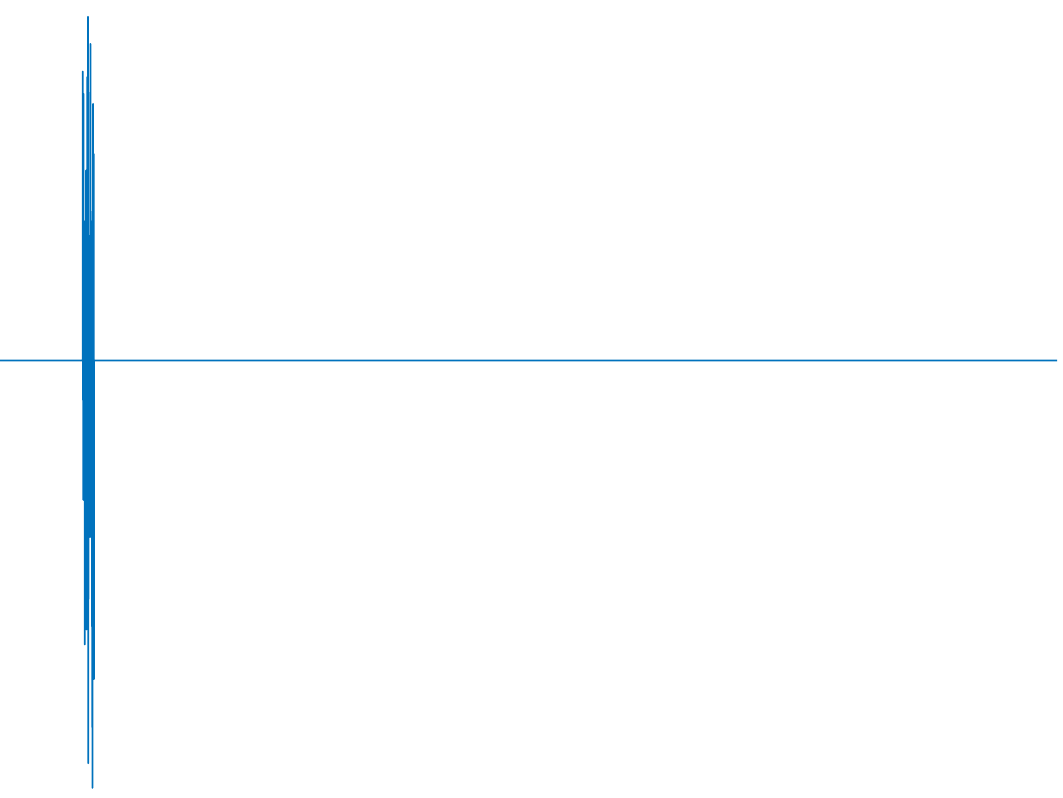}
\end{minipage}\label{fig:jammingsignal}}
\subfigure[$JE=1$kJ]{
\begin{minipage}[b]{0.25\linewidth}
\includegraphics[width=1\linewidth,height=1.9cm]{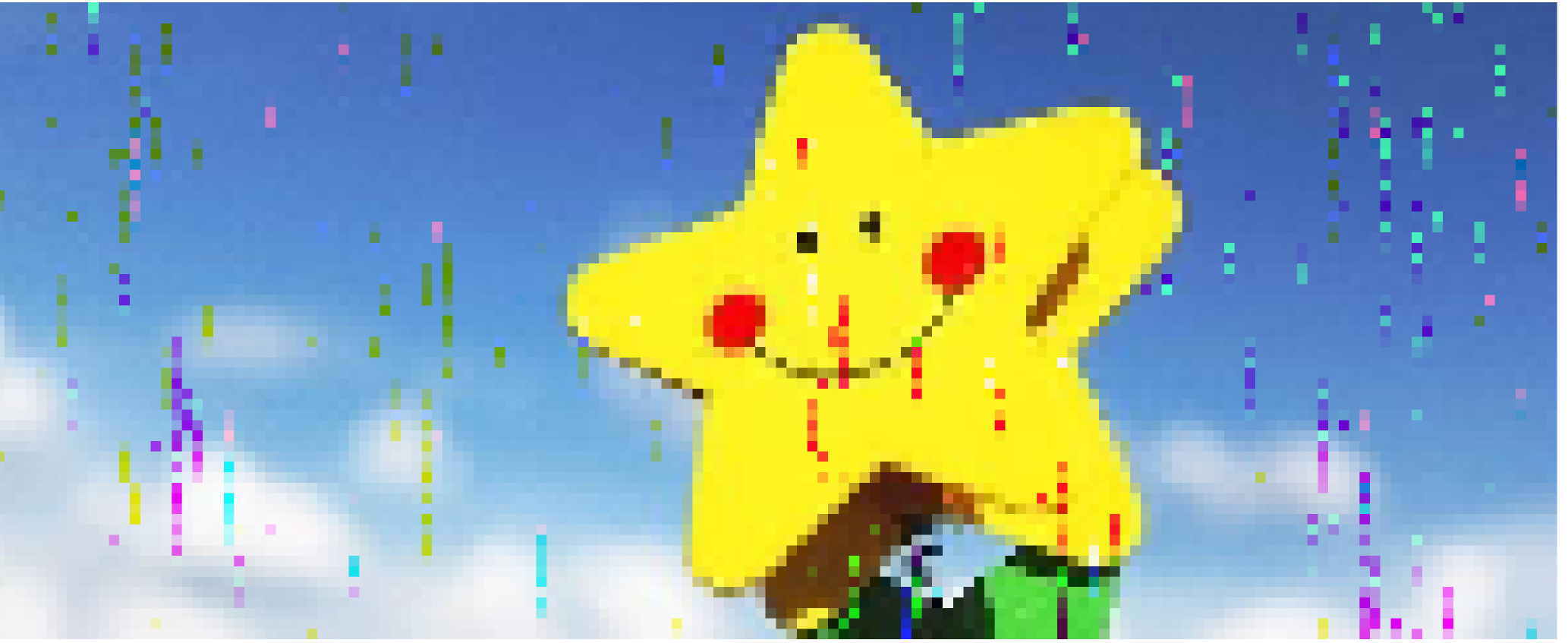}\vspace{4pt}
\includegraphics[width=1\linewidth,height=1.9cm]{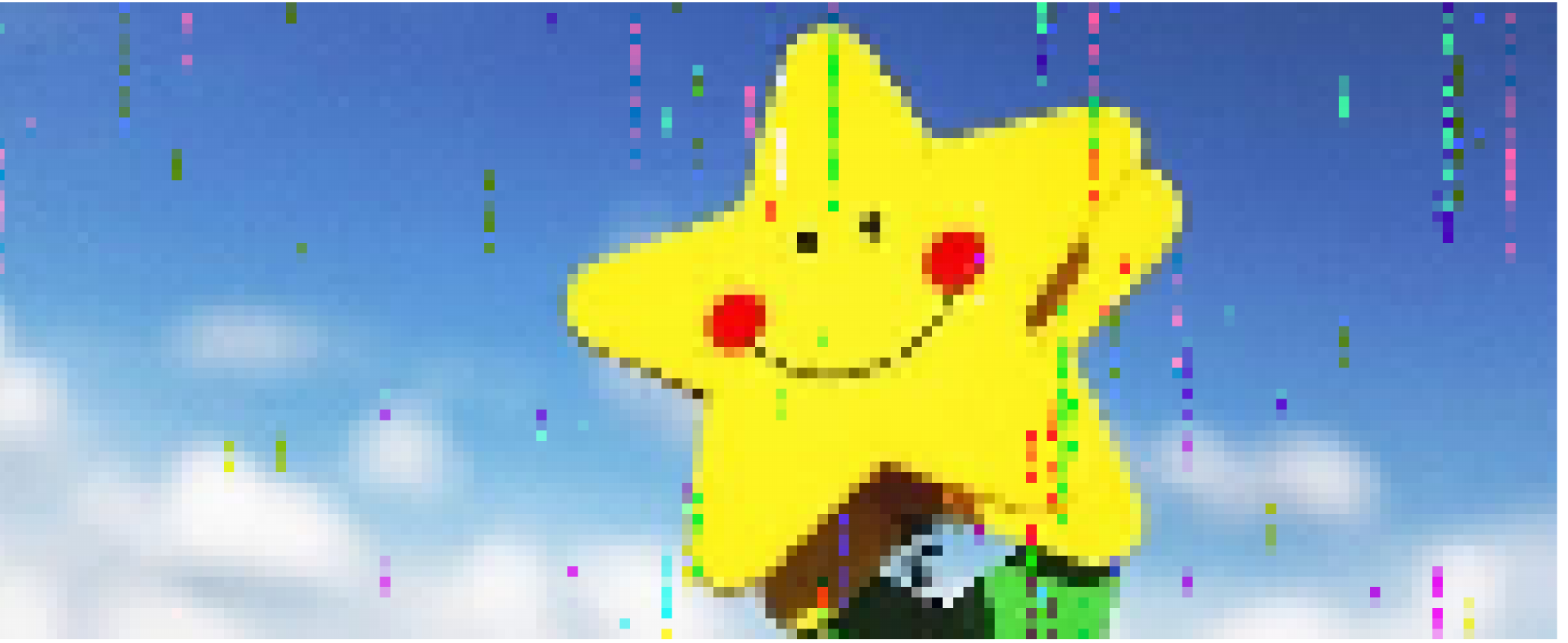}\vspace{4pt}
\includegraphics[width=1\linewidth,height=1.9cm]{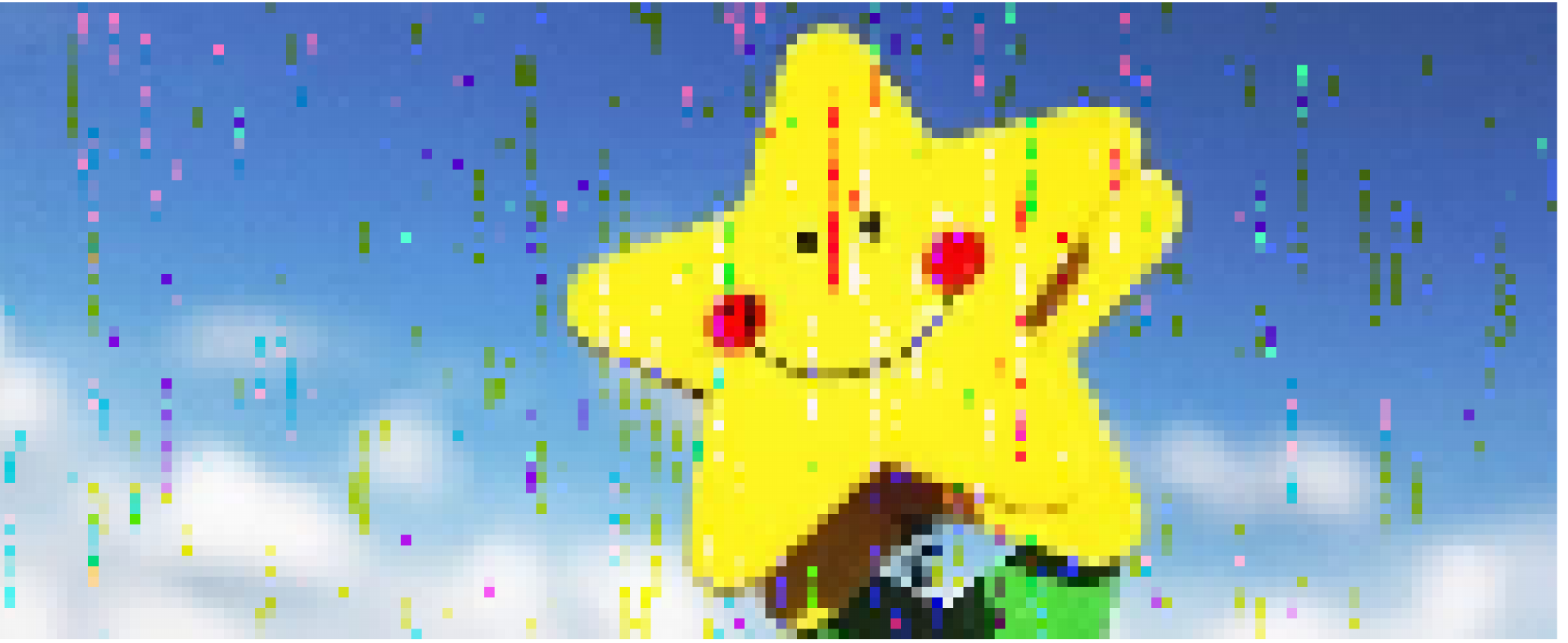}\vspace{4pt}
\includegraphics[width=1\linewidth,height=1.9cm]{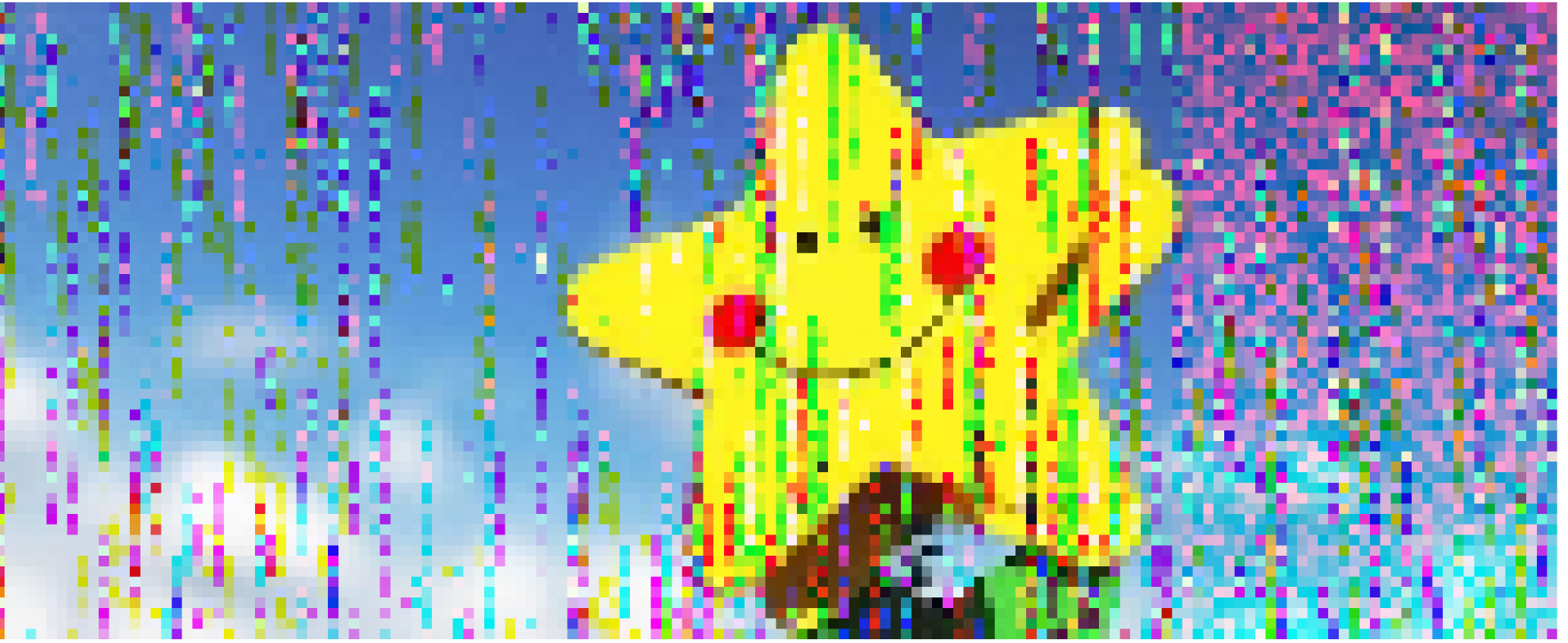}\vspace{4pt}
\includegraphics[width=1\linewidth,height=1.9cm]{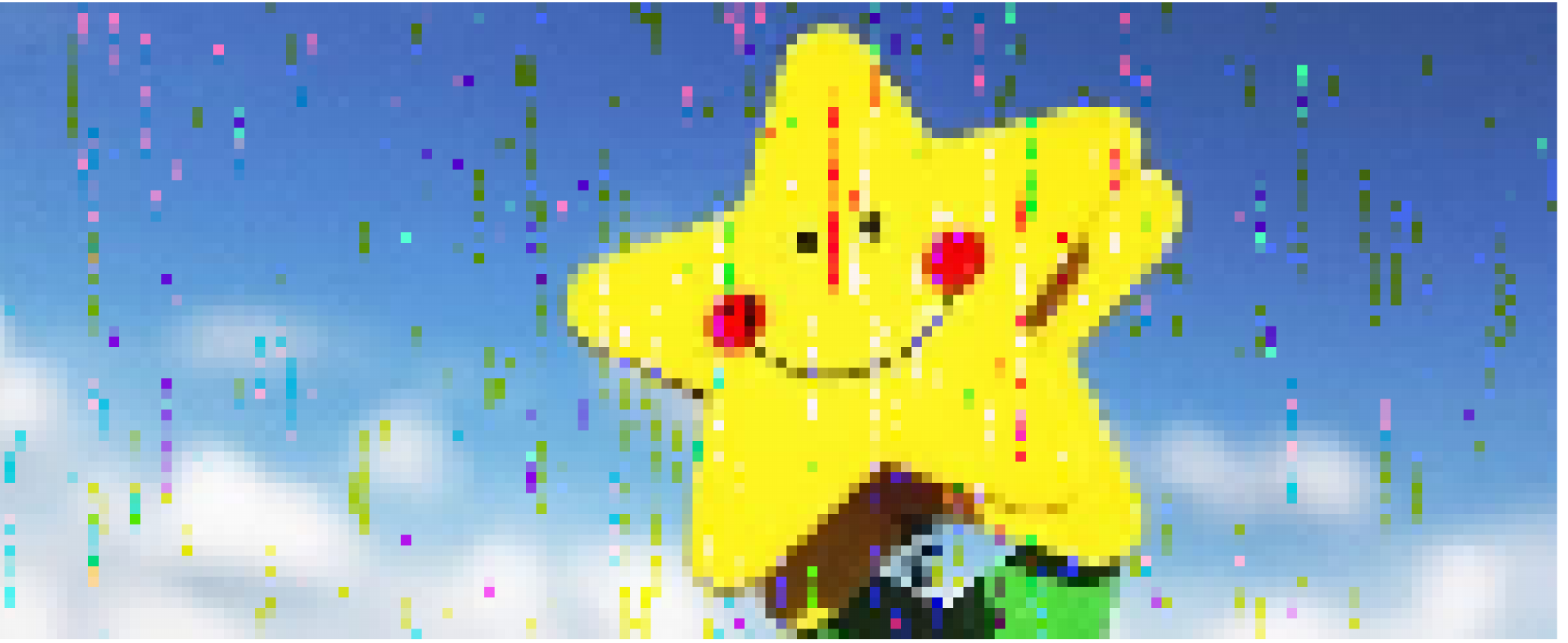}\vspace{4pt}
\includegraphics[width=1\linewidth,height=1.9cm]{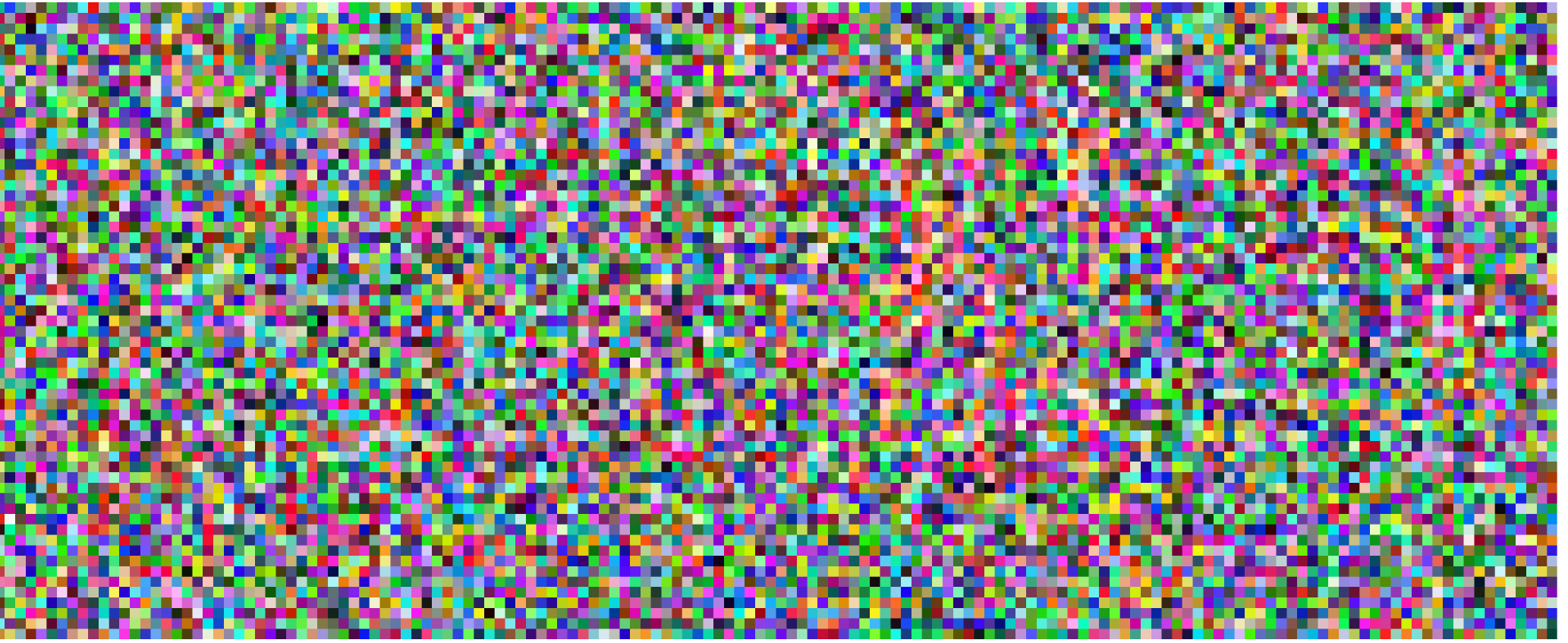}
\end{minipage}\label{fig:JE1}}
\subfigure[$JE=2$kJ]{
\begin{minipage}[b]{0.25\linewidth}
\includegraphics[width=1\linewidth,height=1.9cm]{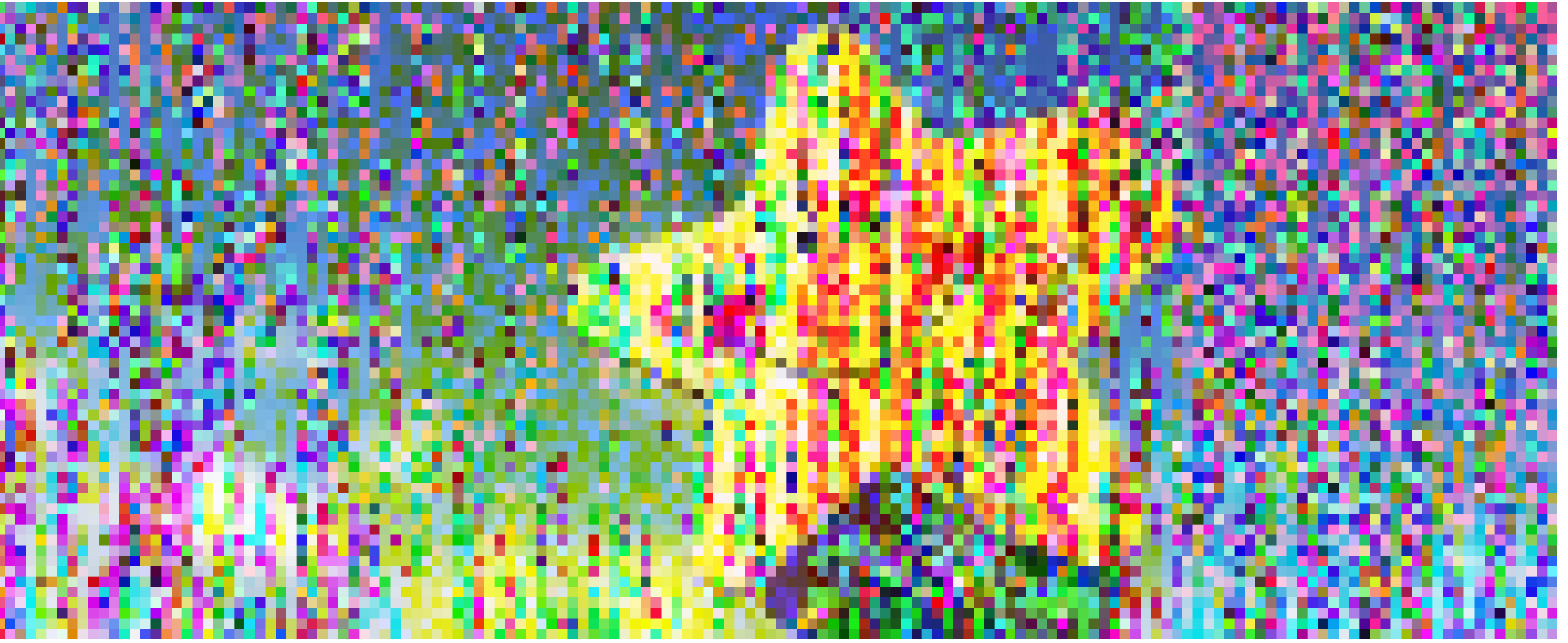}\vspace{4pt}
\includegraphics[width=1\linewidth,height=1.9cm]{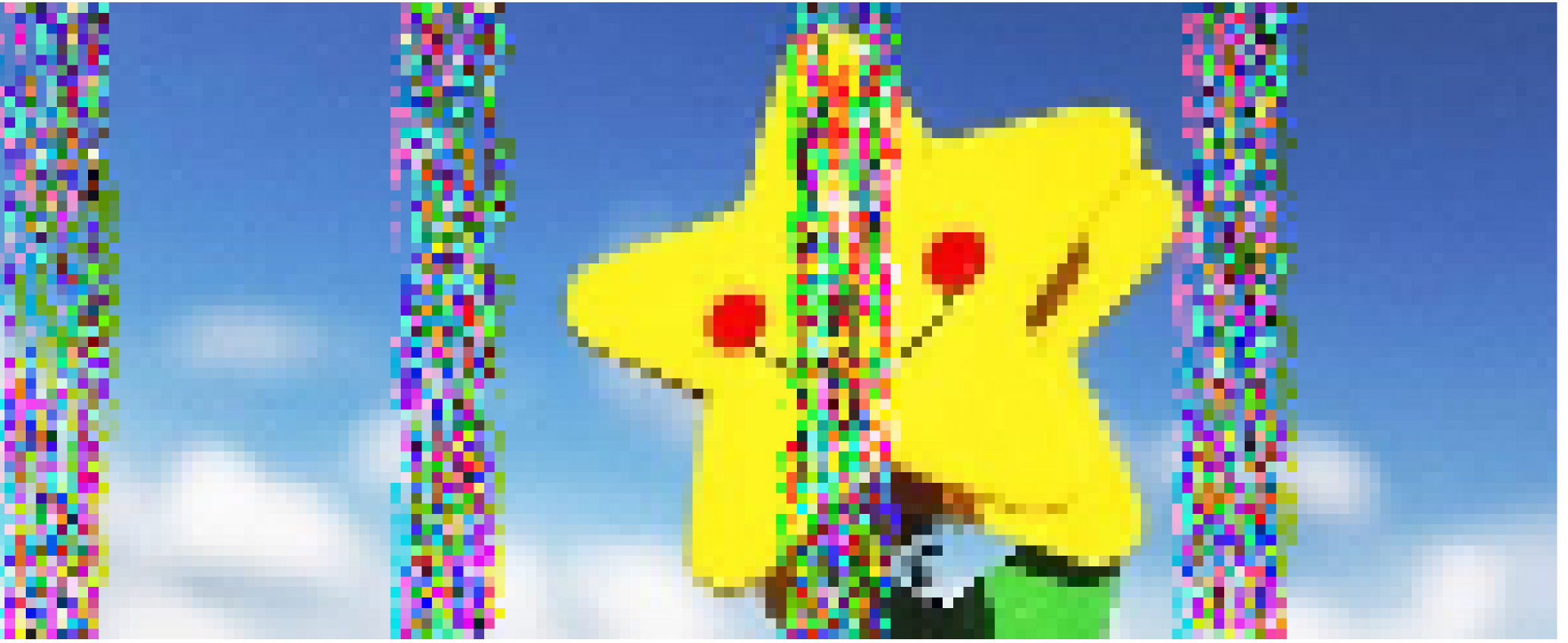}\vspace{4pt}
\includegraphics[width=1\linewidth,height=1.9cm]{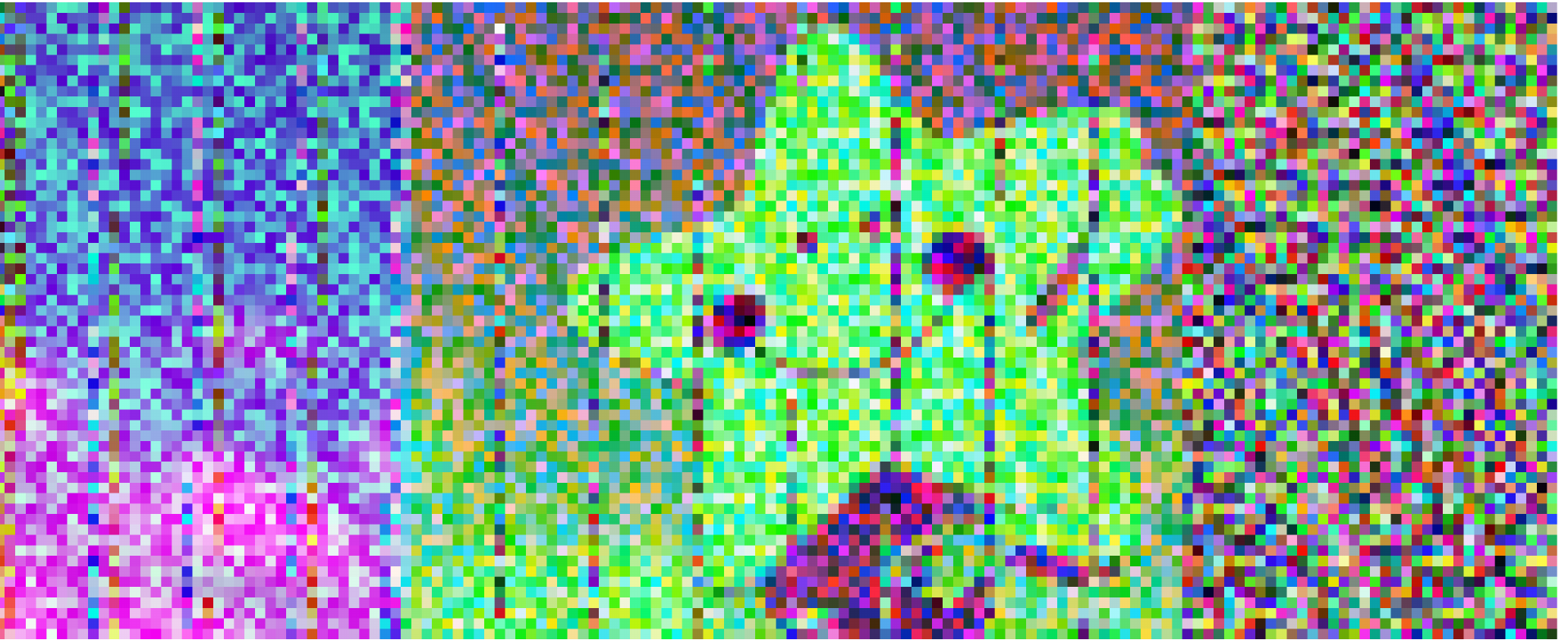}\vspace{4pt}
\includegraphics[width=1\linewidth,height=1.9cm]{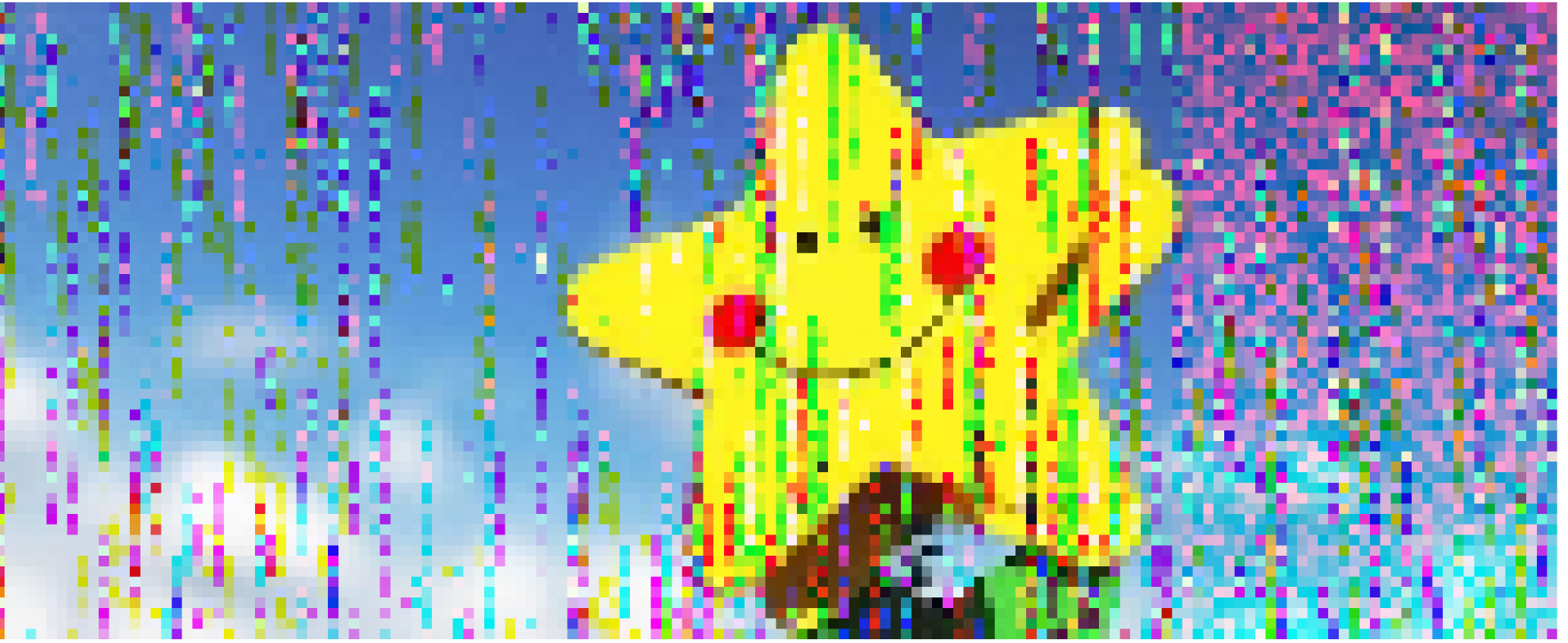}\vspace{4pt}
\includegraphics[width=1\linewidth,height=1.9cm]{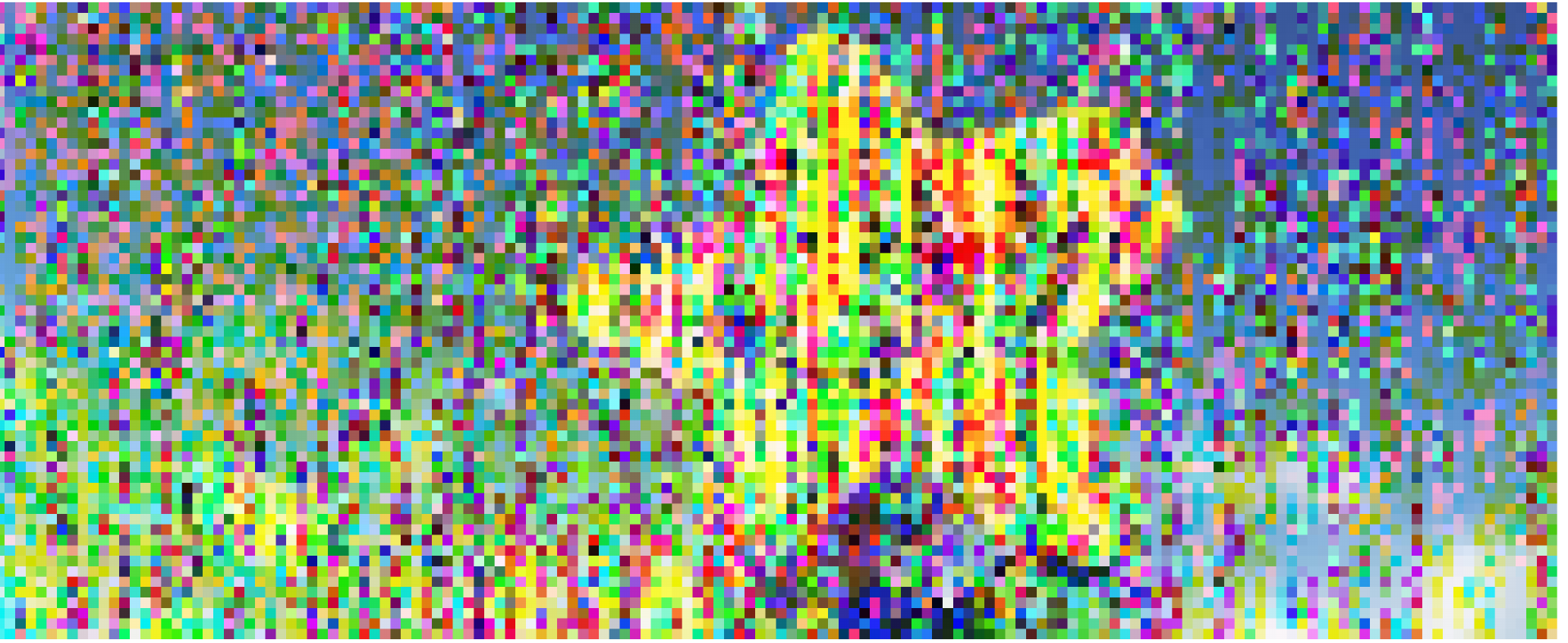}\vspace{4pt}
\includegraphics[width=1\linewidth,height=1.9cm]{figures/1HTLTF.pdf}
\end{minipage}\label{fig:JE2}}
\caption{Illustration of the jamming effects under different jamming energy and different jamming methods.}\label{fig:turesults}
\end{figure*}

\subsection{Jamming Results}
\subsubsection{Received picture of Eve under different IJSs}
The jamming effects under different jamming energy and jamming methods are shown in   Fig.~\ref{fig:turesults}. From top to bottom, Fig.~\ref{fig:jammingsignal} demonstrates the jamming signals transmitted under the continuous jamming scheme (CJS), the PerJDT, the RepJDT, the RanJDT, the RandJFT, and the PerJPT. Note that the demonstration corresponds to the transmission of one data packet, whose repetitions correspond to the whole data transmission process. We can see that the jamming signals are consistent with description in Section \ref{sec:jammingmethod}.

Parallel to different jamming methods in Fig.~\ref{fig:jammingsignal}, the pictures in Fig.~\ref{fig:JE1} illustrate the received results in Eve when the jamming energy is 1kJ. Similarly, Fig.~\ref{fig:JE2} exhibits the received pictures when the jamming energy is 2kJ. There is an intuitive result that the received errors increase with the jamming energy for each jamming method. It is obvious that  periodically jamming the preamble can achieve the best jamming effect under both jamming energy, that is, the legitimate message is completely unavailable to Eve.

\begin{figure*}[htbp]
  \centering
 \subfigure[PerJDT]
{\includegraphics[scale =0.4]{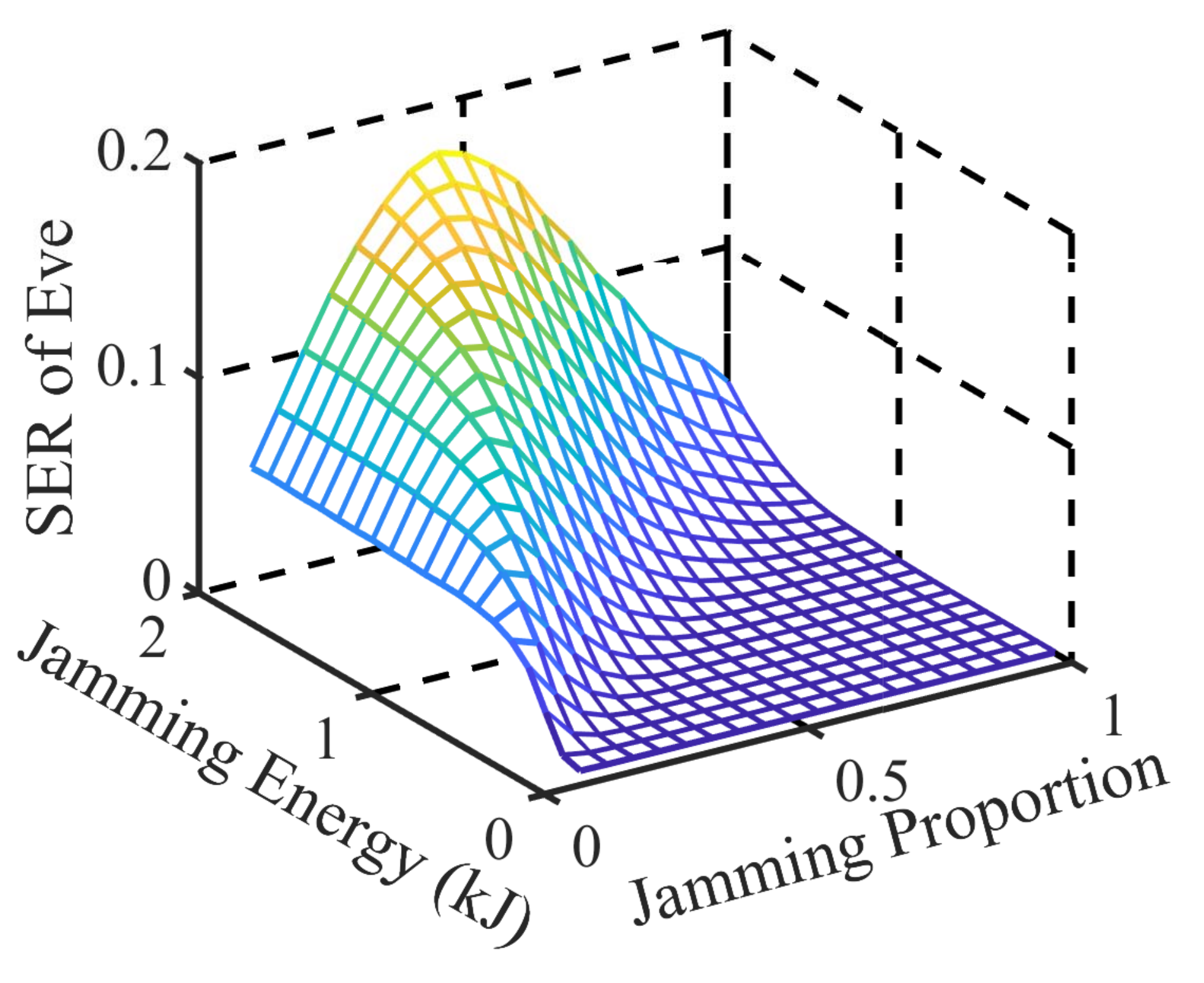}\label{fig:periodBER}}
\subfigure[RepJDT.]
{\includegraphics[scale =0.4]{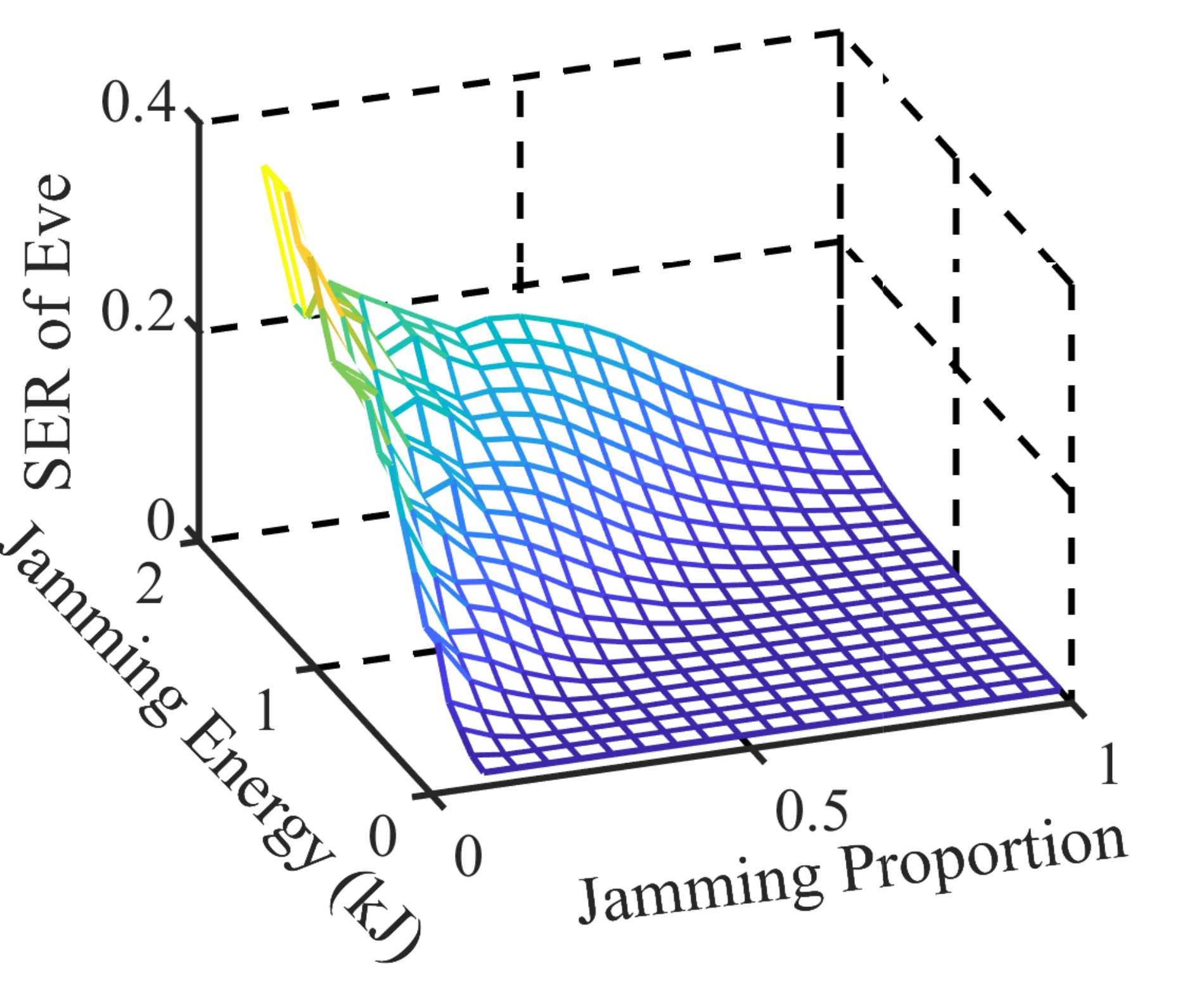}\label{fig:repeateDataBER}}
\subfigure[RanJDT.]
{\includegraphics[scale =0.35]{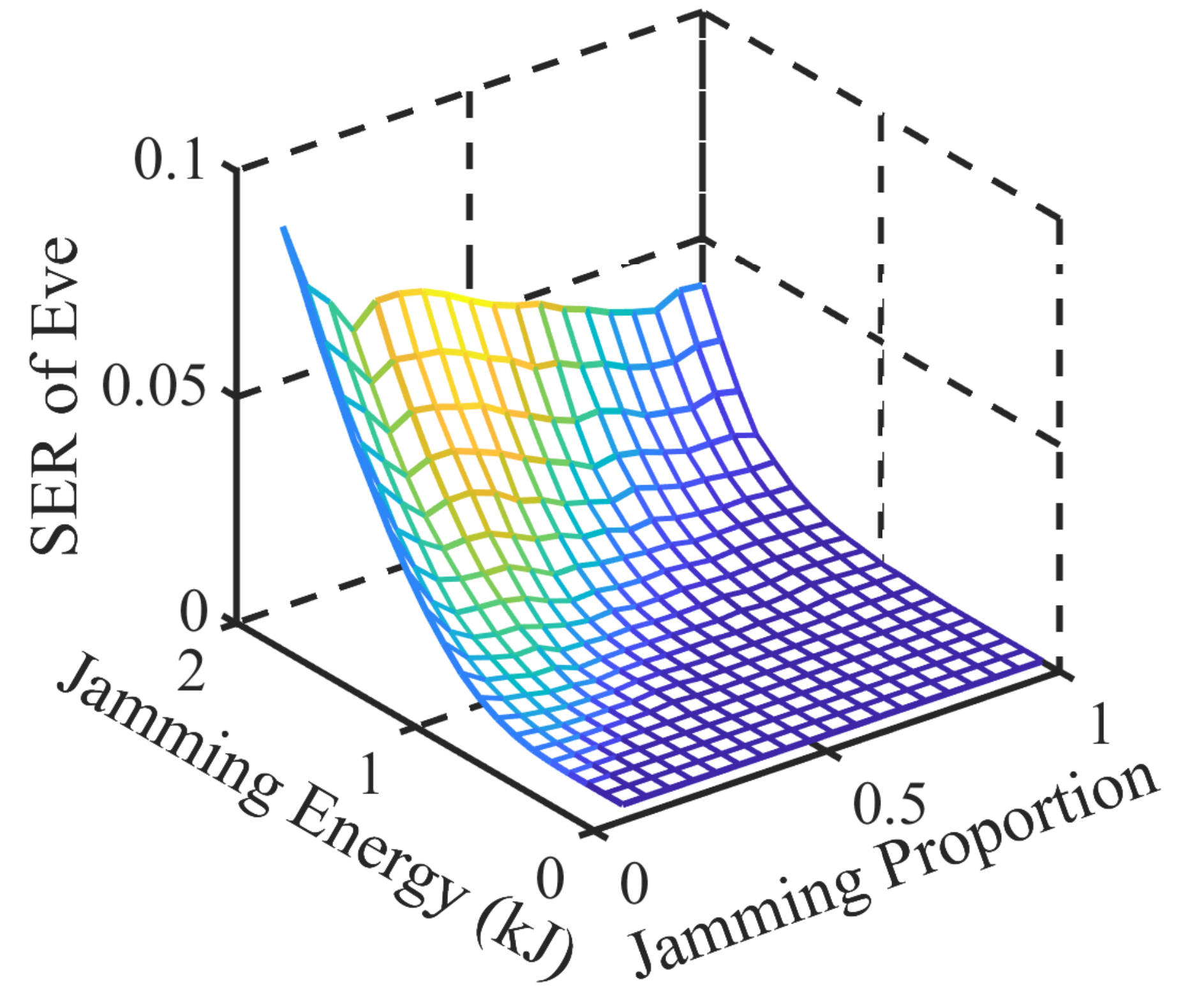}\label{fig:randomDataBER}}
\subfigure[RanJFT.]
{\includegraphics[scale =0.4]{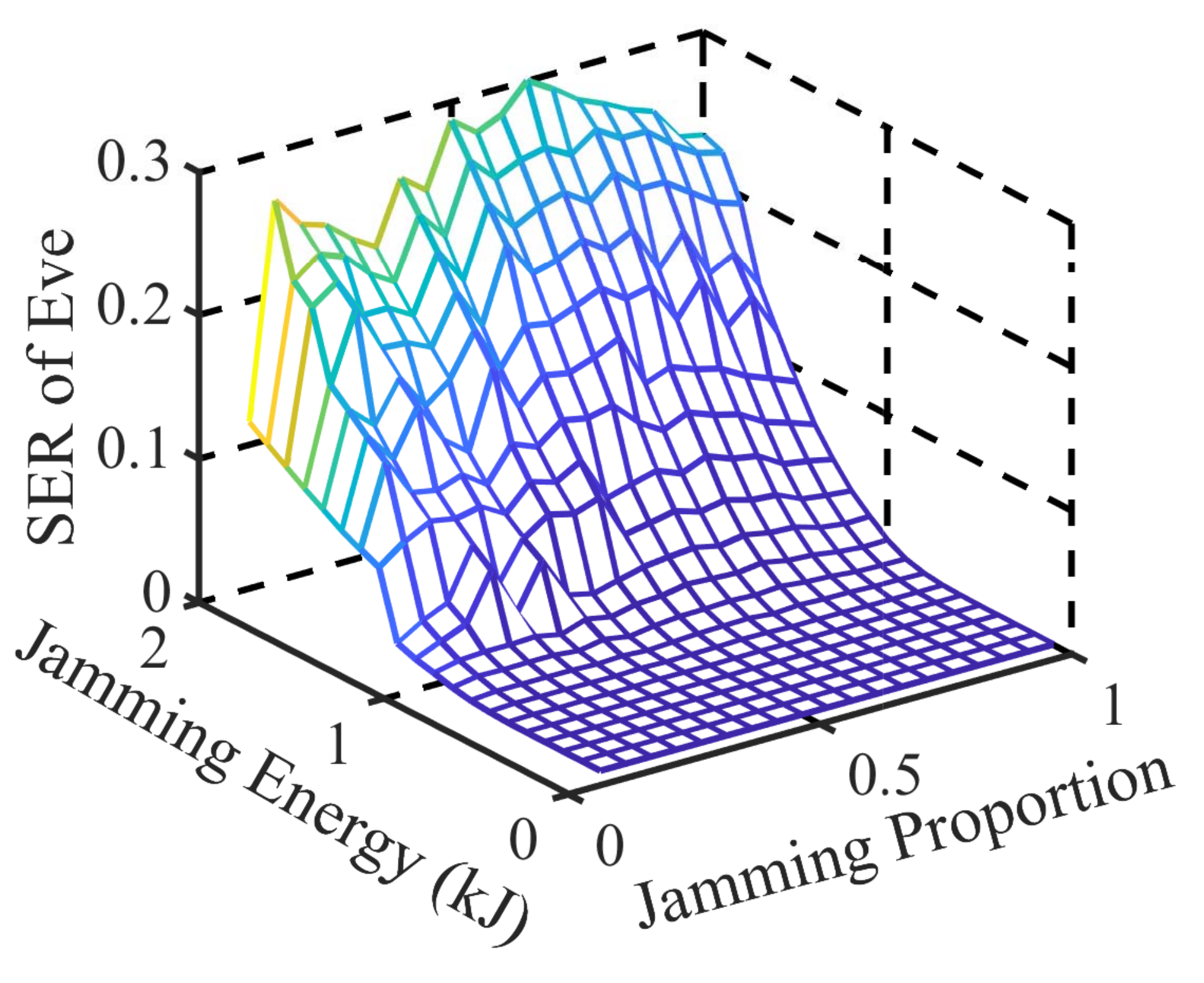}\label{fig:randomPktBER}}
\caption{SER of Eve vs. the jamming energy and the jamming proportion  for some IJSs}\label{fig:BERIJS}
\end{figure*}

\subsubsection{The performance under different jamming energy and jamming proportion}
During the above experiment, we discovered that the numSER of the errors under four IJSs varied with the jamming proportion for a certain jamming energy.  Note that the jamming proportion of the PerJPT depends on the length of the preamble that is fixed. Thus we simulate the SER of Eve versus both the jamming energy and the jamming proportion under four IJSs shown as Fig.~\ref{fig:BERIJS}. In this figure, the SERs are mapped onto the jet color matrix, which ranges between blue and red and passes through cyan and yellow. Specifically, a value of 0 corresponds to the color blue, and a value of 1 refers to the color red. Since the SER is in the range of [0, 0.5], there is no red color in these figures.

From Fig.~\ref{fig:periodBER}-\ref{fig:randomPktBER}, we can see that the SER of Eve for all jamming schemes under a certain jamming proportion increases with the growth of the jamming energy. This is a rational result that a higher jamming energy means a higher influence. As is shown in Fig.~\ref{fig:periodBER}, under a certain jamming energy, the achieved SER for PerJDT first increases then decreases with the increasing jamming proportion. When the jamming proportion is lower than a threshold, the jamming power is high enough to cause errors on the jamming samples, the SER increases with the numSER of the jamming samples. While the jamming proportion increases, the jamming power decreases resulting in a lower SER.

For Fig.~\ref{fig:repeateDataBER}, the RepJDT  is similar to the PerJDT. Specifically, when the jamming proportion is lower, the jamming pulses are narrower and the distances between jamming pulses are longer. Under a certain jamming energy, the error probability is same, but different jamming samples correspond to different transmitted bits. Multiple jamming positions can bring about more bit errors. With the increase of the jamming proportion, the distance between jamming pulses are shortened, which makes this case more like the case of periodically jamming the payload, thus the SER performance is consistent with Fig.~\ref{fig:periodBER}.

Fig.~\ref{fig:randomDataBER} demonstrates the SER performance of RanJDT. We can see that the  overall SER is lower than other three schemes. This is because the jamming pulses are diverse, as shown in Fig.~\ref{fig:jammingsignal}, even the jamming power is higher enough to cause error of the jammed samples, diverse errors cannot mistake a bit. This can also explain the result that the SER under a certain jamming energy, the SER first increases with the jamming proportion. When the jamming proportion goes beyond a threshold, the SER decreases because the jamming power declines much.

While for the case of RanJFT, the SER shown in Fig.~\ref{fig:randomPktBER} still rises with the increasing  jamming energy like other schemes. For a certain jamming energy, the SER vibrates irregularly when the jamming proportion is lower but tends to steadily decrease when the jamming energy goes beyond a threshold. This can also be explained by the different degrees of the dispersion under different jamming proportions like in the above case. Also, this case outperforms the case of RanJDT for same jamming energy and equal jamming proportion, because the probability to jam the preamble in this case is higher than that case.

\subsubsection{SER comparison for different IJSs}
\begin{figure}[!htbp]
  \centering
  \includegraphics[scale=0.35]{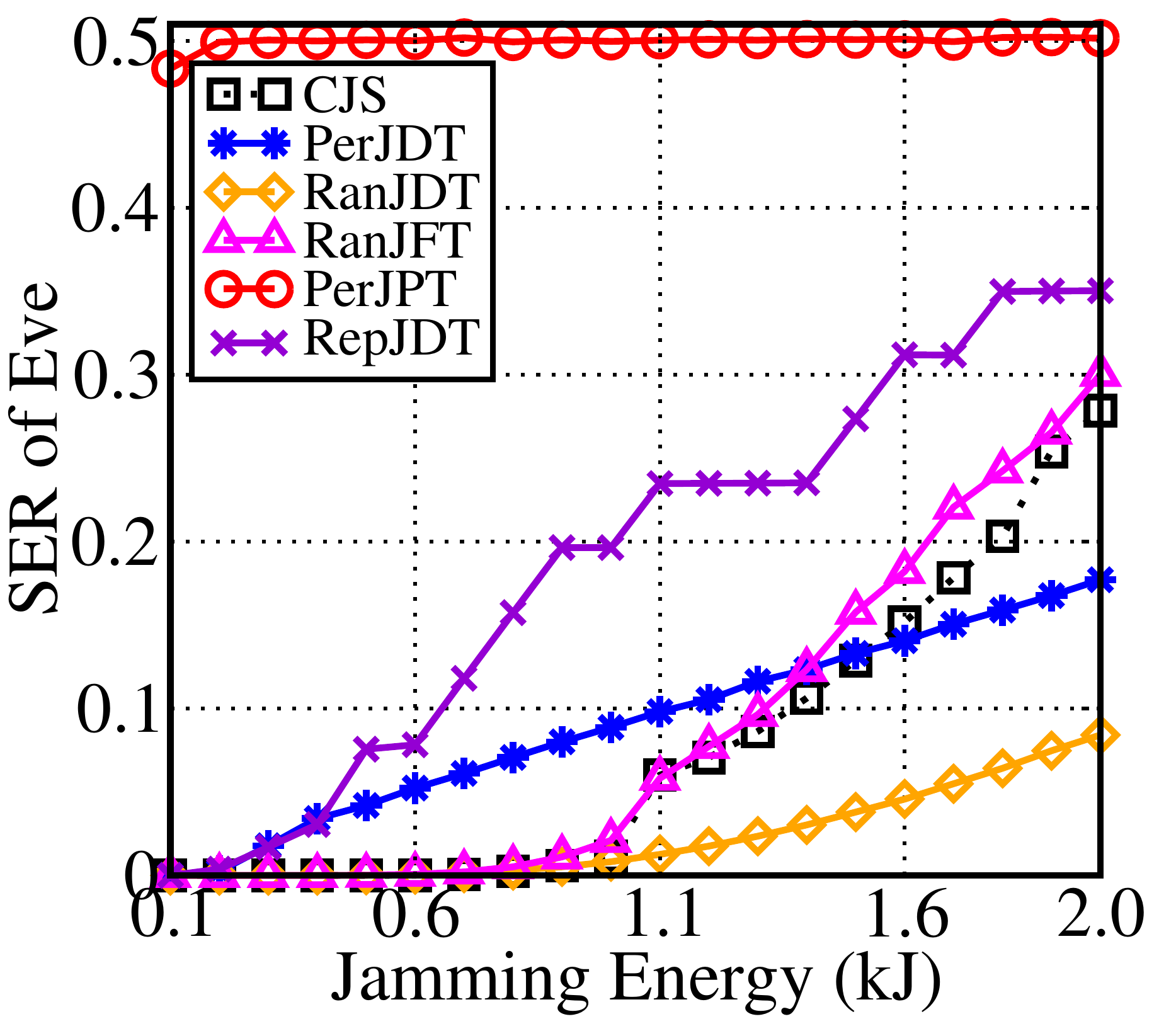}\\
  \caption{Comparison of the best SERs for different jamming schemes under different jamming energy.}\label{fig:comparison}
\end{figure}

For the sake of clearly comparing the performance of all jamming schemes, we plot their best SERs under different jamming energy as illustrated in Fig.~\ref{fig:comparison}. There is no doubt that the growth of the jamming energy contributes to higher SER for all jamming schemes except for the case of preamble-targeted jamming. Under the given range of the jamming energy, jamming the preamble can obtain the maximum SER, i.e., 0.5. Once the HTLTF is jammed, the reference noise level is changed resulting in completely mistaken decoding of the following data part.

For the other jamming schemes, we take the CJS scheme as the reference one. We can see that the jamming effect of RepJDT outperforms the CJS. The reasons are two-folds: i) the RepJDT is a transformed periodically jamming method, thus a jamming pulse can cause continuous errors of a numSER of samples that are matched with a mistaken bit; ii) the dispersion of such jamming pulses in RepJDT can lead to errors of multiple bits. Then, the SER obtained by RanJFT is most close to but a little higher than that of the CJS. Both of them interfere the whole packet especially when the jamming proportion is higher than a threshold (e.g., 0.5), RanJFT is more similar to the CJS but the jamming power is higher than it with the jamming proportion being lower than 1. Next, the jamming effect of PerJDT is first better then worse than that of CJS. This is consistent with the results shown in Fig.~\ref{fig:periodBER}. Finally, the worst jamming effect is obtained by RanJDT due to the dispersion of the jamming signals.

In summary, the application of the proposed schemes can be concluded as follows. When we have little energy, we can select the scheme that can cause a larger SER. Correspondingly, when we have a given SER, we can choose the one than can consume lower energy. Further, we can also jointly consider the difficulty of generating the jamming pulses, the synchronization, and the inference control problems.

\section{Conclusion}
\label{sec:conclusion}
In this paper, we investigated the intermittent jamming schemes for secrecy and energy efficiency enhancement in the energy-constraint networks. Based on the WLAN frame format, we proposed five intermittent jamming schemes, including the PerJPT, PerJDT, RepJDT, RanJDT, and the RanJFT, according to the jamming positions and jamming methods. We discussed their applicability according to requirements on the synchronization, the signal detection, and the interference cancellation at Bob. Simulation experiments were conducted under different available jamming energy and jamming proportions and demonstrated the advantages of the proposed IJS over the traditional CJS in increasing the Eve's BER. As a part of our future work, we consider the coordinate work of multiple intermittent jammers to enhance the security of the energy-constraint dense cooperative networks.

\section*{Acknowledgment}
This work is supported by the National Natural Foundation of China (Grant No. 61871023 and 61931001), the Youth Program of National Natural Science Foundation of China
(Grant No. 61802274), and the Natural Science Found for Colleges and Universities in Jiangsu Province (18KJB510044).

\end{document}